\documentclass[manuscript]{aastex62} 
\usepackage{amsmath}
\begin{document}

\title{The Effect of Far-infrared Radiation on the Hyperfine Anomaly of the OH 18 cm Transition}


\author{Yuji Ebisawa}
\affiliation{Department of Physics, The University of Tokyo, Hongo, Bunkyo-ku, Tokyo 113-0033, Japan}
\author{Nami Sakai}
\affiliation{RIKEN, 2-1 Hirosawa, Wako, Saitama 351-0198, Japan}
\author{Karl M. Menten}
\affiliation{Max-Planck-Institut f\"{u}r Radioastronome, Auf dem H\"{u}gel 69, D-53121 Bonn, Germany}
\author{Satoshi Yamamoto}
\affiliation{Department of Physics, The University of Tokyo, Hongo, Bunkyo-ku, Tokyo 113-0033, Japan}
\begin{abstract}
    We present observations of the four hyperfine structure components of the OH 18 cm transition (1612, 1665, 1667 and 1720 MHz) toward Heiles Cloud 2 (HCL2) in Taurus and the dense cores L183 and L169.
    Toward the peculiar straight structure in the northern part of HCL2, the 1720 MHz line shows absorption against the cosmic microwave background at a velocity of $\sim$ 5.5 km s$^{-1}$,
    whereas the 1612 MHz line shows brighter emission than expected in local thermodynamic equilibrium (LTE).
    Such an intensity anomaly is also observed toward TMC-1 (CP), L183, and L169,
    where the 1612 MHz line is stronger and the 1720 MHz line is weaker than those expected under LTE.
    We conduct statistical equilibrium calculations considering the effect of far-infrared (FIR) radiation from surrounding clouds
    and find that the absorption feature of the 1720 MHz line can be reproduced by the following conditions:
    gas kinetic temperature lower than about 30 K, OH column density
    higher than 10$^{15}$ cm$^{-2}$, H$_2$ density lower than 10$^3$ cm$^{-3}$,
    and an ortho-to-para ratio of H$_2$ lower than 1.
    Therefore, the absorption feature of the 1720 MHz line is considered to trace
    relatively cold and dense gas that is surrounded by a warm envelope consisting of CO-dark molecular gas.
\end{abstract}
\section{INTRODUCTION} \label{sec:Introduction}
\begin{figure}[htbp] 
  \centering
  \epsscale{0.5}
  \plotone{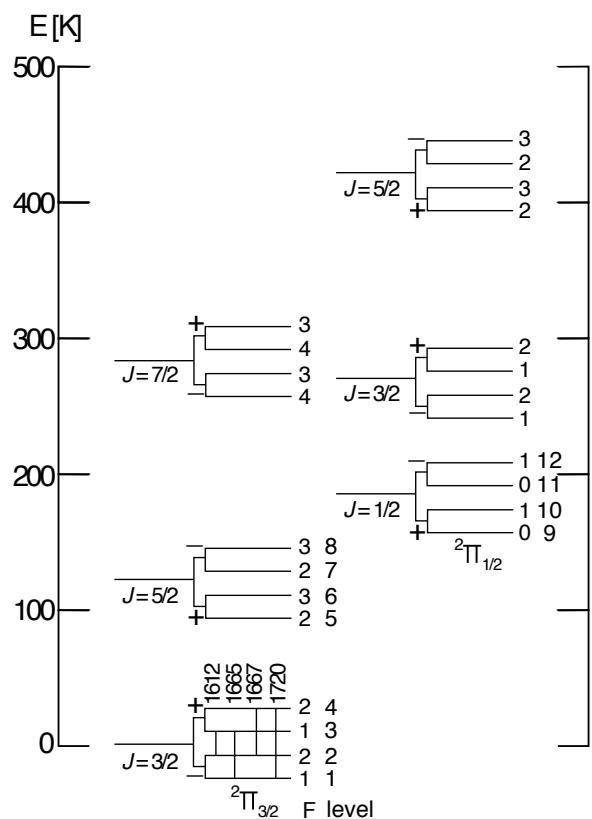}
  \caption{%
    Rotational energy level structure of OH.
    The display of the fine and hyperfine structure levels is schematic,
    and the separations of the split energy levels are not drawn to scale.
    \label{fig:OH_energy}}
\end{figure}

    The 18 cm transition of the hydroxyl radical (OH) has been a unique probe of diffuse and dense interstellar clouds, late-type stars, and external galaxies \citep[e.g.][]{Harju2000, Caswell2004, Hoffman2005, Wolak2012, Darling2002}.
    It is the $\Lambda$-type doubling transition in the ground rotational state ($J=3/2$ in $^2\Pi_{3/2}$).
    Due to the nuclear spin of the proton, the OH 18 cm transition is split into four hyperfine component lines, as shown in Figure $\ref{fig:OH_energy}$.
    In the interstellar medium, the relative population of the four hyperfine structure (hfs) levels often deviates from the values attained in local thermodynamic equilibrium (LTE).
    This is caused by collisional/radiative excitation to the
    rotationally excited states
    and subsequent de-excitation to the ground rotational state \citep{Elitzur1976b, Langevelde1995}.
    Hence the relative intensities of the four hfs component lines at 1612, 1665, 1667, and 1720 MHz differ from the intrinsic intensity ratio of $1:5:9:1$.
    In extreme cases, some of these hfs component lines display maser emission \citep{Gundermann1965, Weaver1965, Cohen1995}.
    Moreover, the 1612 and 1720 MHz lines (called satellite lines) may appear in absorption and emission, respectively, toward sources with bright continuum emission.
    This is known as so-called ``conjugate'' behavior \citep{Langevelde1995, Weisberg2005, Kanekar2004}.

    Recently, we have shown that the above anomaly of hyperfine intensities can be used as
    an effective thermometer of molecular clouds over a wide range of H$_2$ density \citep{Ebisawa2015}.
    According to our statistical equilibrium calculations for OH, the 1612 MHz line is observed in absorption against the cosmic microwave background (CMB) for the gas kinetic temperatures of 40 K or higher.
    The 1720 MHz line intensity shows the conjugate behavior to the 1612 MHz line, and its intensity is higher than that expected under LTE conditions.
    This behavior is almost independent of the H$_2$ density from 10$^2$ to 10$^6$ cm$^{-3}$.
    Indeed, the anomaly of the 1612 MHz and 1720 MHz lines was observed with the Effelsberg 100 m radio telescope toward the east part of Heiles Cloud 2 (HCL2E), the dark cloud L183, and the $\rho$-Ophiuchi molecular cloud.
    This allowed accurate determinations of gas kinetic temperatures.
    For instance, the temperature is evaluated to be 53 $\pm$ 1 K toward HCL2E.
    In $\rho$-Ophiuchi, the temperature tends to be lower for more distant positions from the nearby illuminating star HD147889.
    Thus, the OH 18 cm transition can be used to trace a new class of warm molecular gas surrounding a molecular cloud
    that is not well traced by emission of CO and its isotopologues
    (i.e., so-called CO-dark molecular gas).

    In the course of our survey of the OH 18 cm transition toward various molecular clouds,
    we noticed another type of the hyperfine anomaly.
    Toward the cold dark cloud L183, the 1612 MHz line is brighter than the 1720 MHz
    \replaced{line, a behavior that}{{line. This behavior}} is opposite to the above hyperfine anomaly.
    The same trend was pointed out in the north part of TMC-1 ridge by \citet{Harju2000}.
    Most notably, we have found an absorption feature in the 1720 MHz line and bright emission in the 1612 MHz line toward the north end of HCL2.
    Recently, \citet{Xu2016a} also reported the same anomaly in the northern part of HCL2 with the Arecibo 300 m telescope.
    Such an anomaly cannot be reproduced by our statistical equilibrium calculations mentioned above.
    It is predicted to be caused by far-infrared (FIR) pumping of the OH molecules to the second rotationally excited level,
    as discussed by \citet{Harju2000} and \citet{Elitzur1976a}.
    For a comprehensive understanding of the hyperfine anomaly of the OH 18 cm transitions based on the non-LTE excitation calculation,
    here we investigate the anomaly by incorporating the effect of the far-infrared pumping.
\section{OBSERVATIONS} \label{sec:Observation}
    We observed four positions in Heiles Cloud 2 (HCL2) (Figure $\ref{fig:HCL2}$) and the cold dark cloud L183/L169 (Figure $\ref{fig:L183}$) in the OH 18 cm transition
    with the Effelsberg 100 m telescope of the Max-Planck-Institut f{\" u}r Radioastronomie (MPIfR) in 2013 and 2016 October.
    For our observations, the 18 cm/21 cm prime focus receiver was used,
    whose system noise temperature was about 20 K.
    The full-width-half-maximum beam size is 8.2$'$.
    The telescope pointing was maintained to be better than 20$''$, by observing nearby continuum sources.
    The facility Fast Fourier Transform Spectrometer (FFTS) was employed as the back end;
    it has an instantaneous bandwidth of 100 MHz with a frequency resolution of 3.5 kHz.
    This corresponds to a velocity resolution of 0.56 km s$^{-1}$ at 1667 MHz.
    We observed the four hfs components of the OH 18 cm transition with two frequency settings,
    where the 1612 MHz and 1720 MHz lines were separately observed.
    The 1665 MHz and 1667 MHz lines were observed in both frequency settings.
    We used the frequency-switching mode with the frequency offset of 0.1 MHz.
    We hereafter use main beam brightness temperature ($T_{MB}$) as an intensity scale,
    which is calculated from the antenna temperature assuming a main beam efficiency of 0.65.
    The \replaced{RMS}{{rms}} (root-mean-square) noise levels of the observed spectra are about 8--16 mK.
\section{RESULTS} \label{sec:Result}
\begin{figure}[htbp] 
  \centering
  \epsscale{0.8}
  \plotone{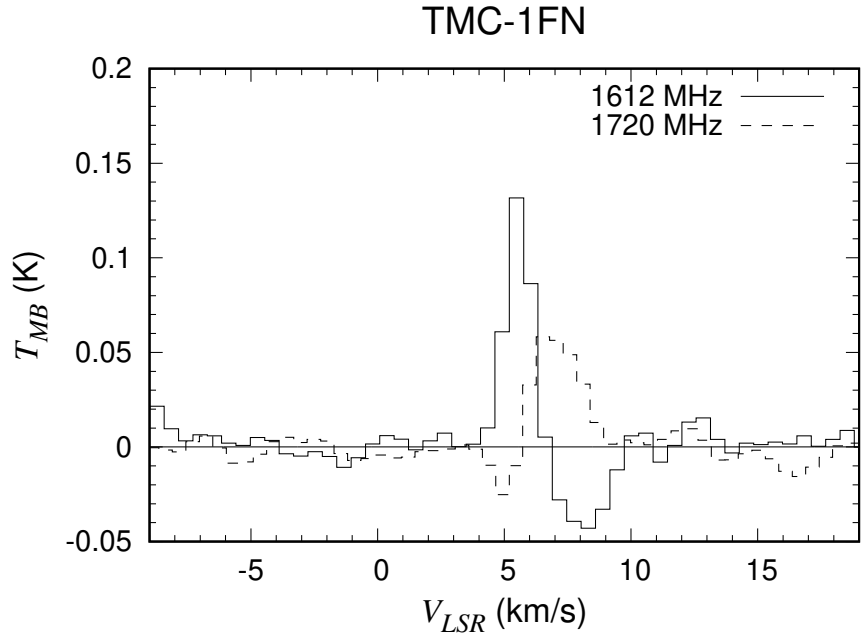}
  \caption{
    The 1612 MHz (solid) and the 1720 MHz (dashed) line spectra observed toward TMC-1FN.
    \label{fig:sTMC1FN_overlay}
  }
\end{figure}

    Figure $\ref{fig:sTMC1FN}$ shows the spectra of the OH 18 cm transition observed on the straight structure in the north part of HCL2 \citep{Narayanan2008, Goldsmith2010}, which we call TMC-1FN (Figure $\ref{fig:HCL2}$).
    These spectra are best explained by three Gaussian profiles centered on $V_{LSR}$=5.5 km s$^{-1}$, 6.5 km s$^{-1}$, and 8.2 km s$^{-1}$, respectively.
    The line parameters are summarized in Table $\ref{table:linepara}$.
    The 1612 MHz line appears in absorption in the red-shifted component ($V_{LSR}$=8.2 km s$^{-1}$), whereas it shows emission in the blue-shifted component ($V_{LSR}$=5.5 km s$^{-1}$).
    It is very faint in the central component ($V_{LSR}$=6.5 km s$^{-1}$).
    The absorption feature of the red-shifted component traces warm gas with a gas kinetic temperature higher than 40 K, according to our statistical equilibrium calculations \citep{Ebisawa2015}.
    On the other hand, {\it the absorption feature of the 1720 MHz line} is observed in the blue-shifted component.
    The conjugate behavior of these satellite lines in the two velocity components is clearly seen in Figure $\ref{fig:sTMC1FN_overlay}$,
    in which the spectra of the 1612 MHz and the 1720 MHz lines are overlaid with each other.
    The 1612 MHz and 1720 MHz lines show absorption and bright emission, respectively, in the red-shifted component, whereas they show the opposite in the blue-shifted component.
    We performed additional statistical equilibrium calculations \citep{Ebisawa2015}, but the 1720 MHz line absorption could not be reproduced
    in spite of an extensive parameter search.
    Although \citet{Xu2016a} reported the absorption feature of the 1720 MHz line on the straight structure of HCL2 located in the northwest of TMC-1FN,
    they could not reproduce it with non-LTE radiative transfer model calculation, either.

\begin{figure}[htbp] 
  \centering
  \epsscale{0.5}
  \plotone{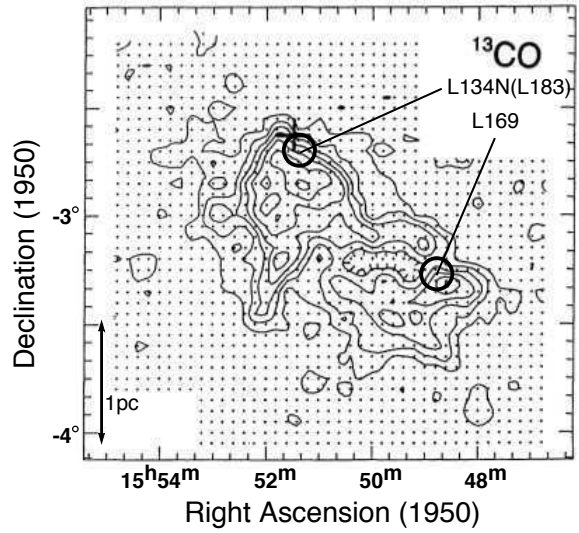}
  \caption{%
    Contours show the integrated intensity map of $^{13}$CO ($J$=1-0) toward L183 reported by \citet{Laureijs1995}.
    The two circles represent the observed position in OH.
    The cross marks the position of the NH$_3$ core of L183 (L134N) indicated by \citet{Laureijs1995}.
    \label{fig:L183}}
\end{figure}

\begin{figure}[htbp] 
  \centering
  \epsscale{1.0}
  \plotone{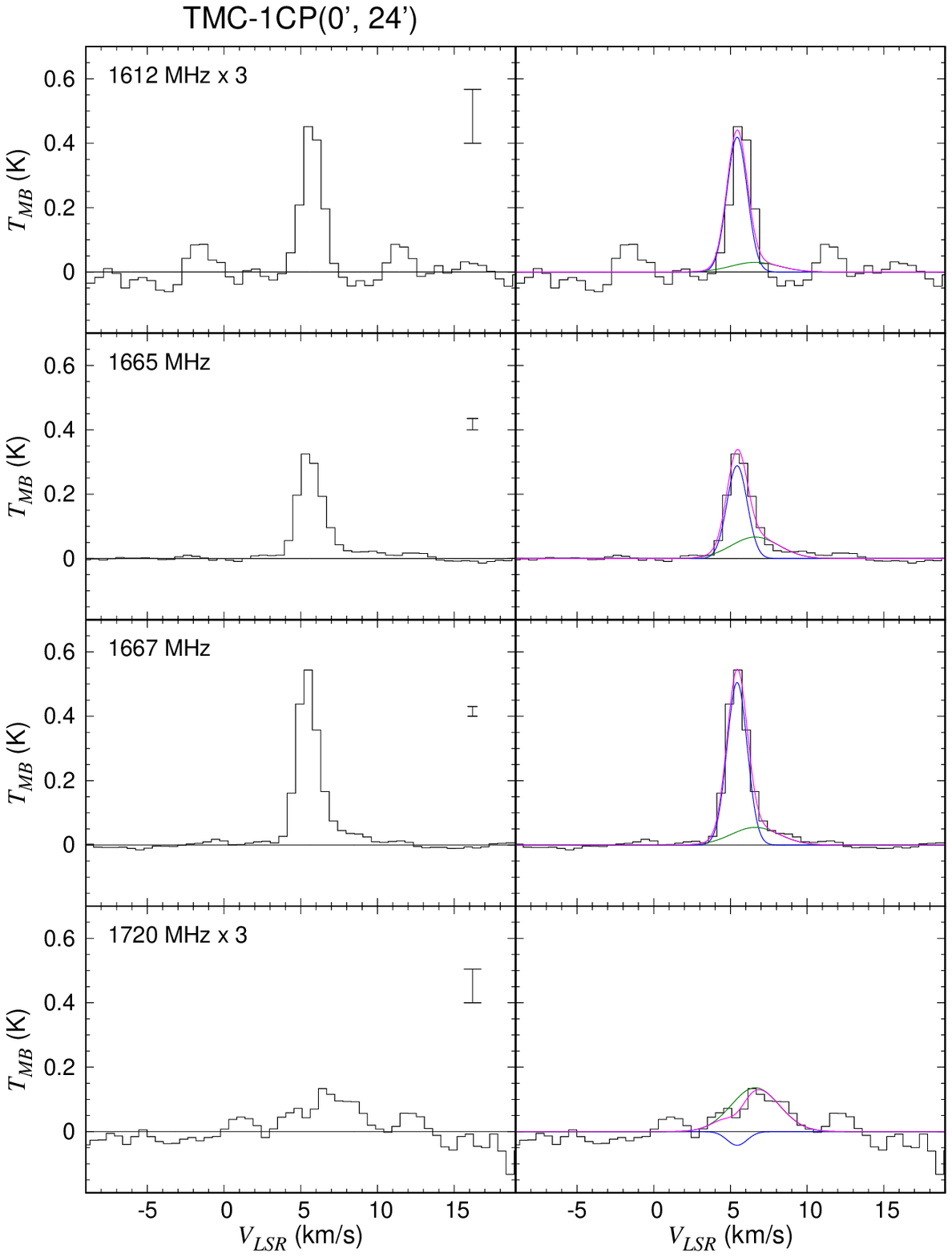}
  \caption{
    (Left) Observed spectra of the OH 18 cm transition toward 24$'$ north from TMC-1(CP).
    Three times the rms noise of each spectrum is represented in the top right.
    (Right) The best-fit Gaussian profiles obtained
    by assuming two velocity components are shown by blue and red lines,
    whose central velocities are 5.4 and 6.6 km s$^{-1}$, respectively.
    Magenta lines \replaced{shows}{{show}} the sum of these three velocity components.
    \label{fig:sTMC1CP_0_24}
  }
\end{figure}

\begin{figure}[htbp] 
  \centering
  \epsscale{1.0}
  \plotone{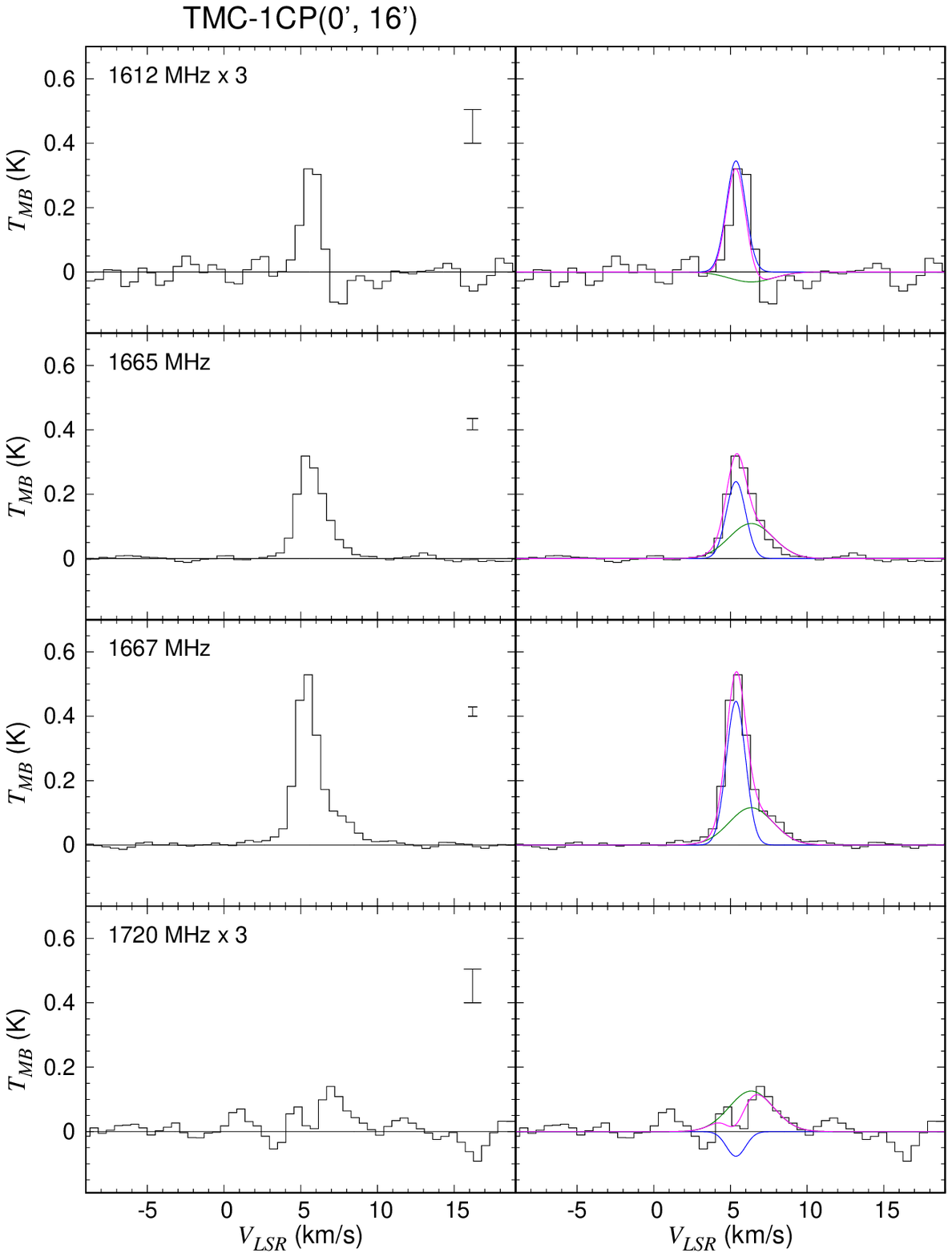}
  \caption{
    (Left) Observed spectra of the OH 18 cm transition toward 16$'$ north from TMC-1(CP).
    Three times the rms noise of each spectrum is represented in the top right.
    (Right) The best-fit Gaussian profiles obtained
    by assuming two velocity components are shown by blue and red lines,
    whose central velocities are 5.4 and 6.4 km s$^{-1}$, respectively.
    Magenta lines \replaced{shows}{{show}} the sum of these three velocity components.
    \label{fig:sTMC1CP_0_16}
  }
\end{figure}

\begin{figure}[htbp] 
  \centering
  \epsscale{1.0}
  \plotone{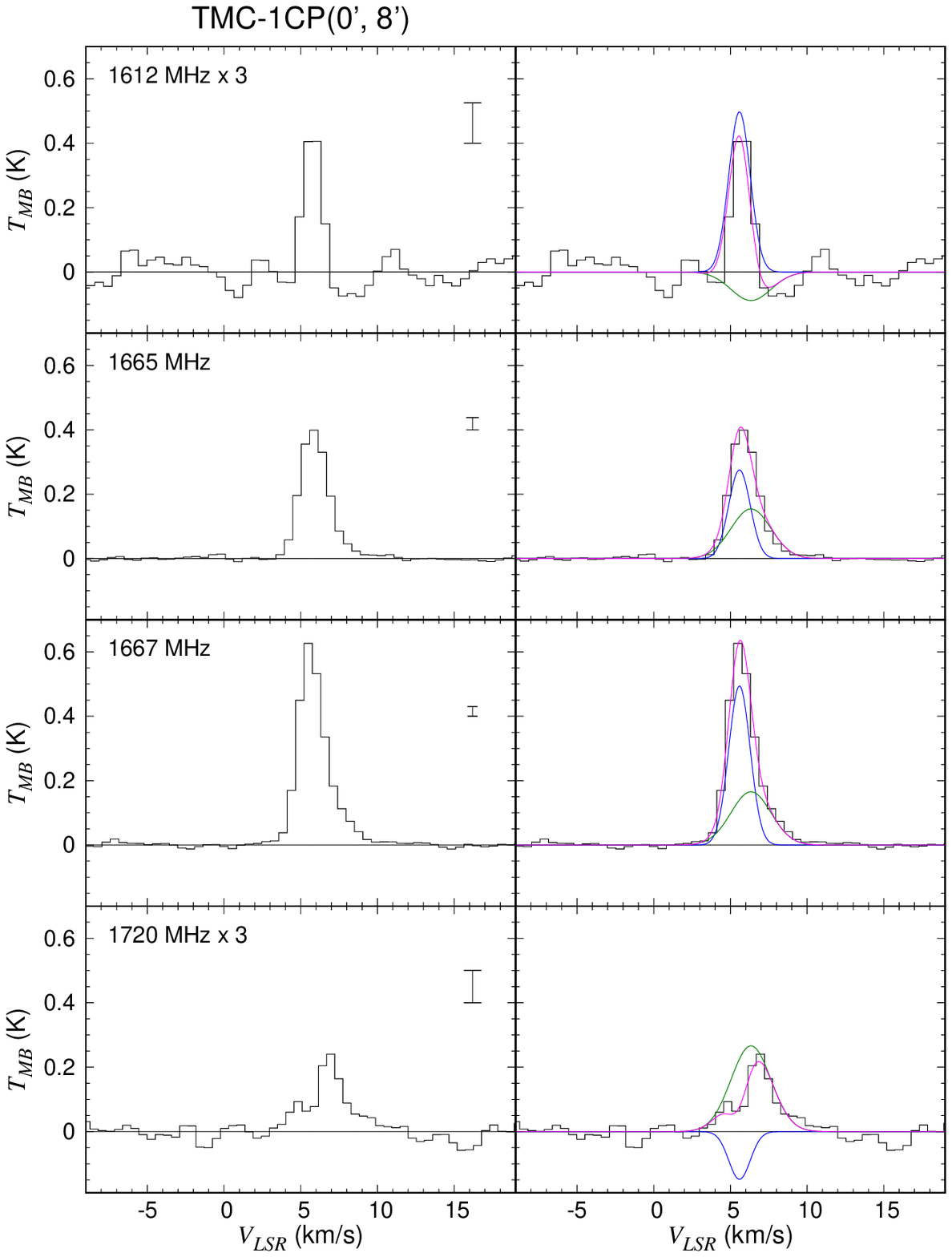}
  \caption{
    (Left) Observed spectra of the OH 18 cm transition toward 8$'$ north from TMC-1(CP).
    Three times the rms noise of each spectrum is represented in the top right.
    (Right) The best-fit Gaussian profiles obtained
    by assuming two velocity components are shown by blue and red lines,
    whose central velocities are 5.6 and 6.3 km s$^{-1}$, respectively.
    Magenta lines \replaced{shows}{{show}} the sum of these three velocity components.
    \label{fig:sTMC1CP_0_8}
  }
\end{figure}

\begin{figure}[htbp] 
  \centering
  \epsscale{1.0}
  \plotone{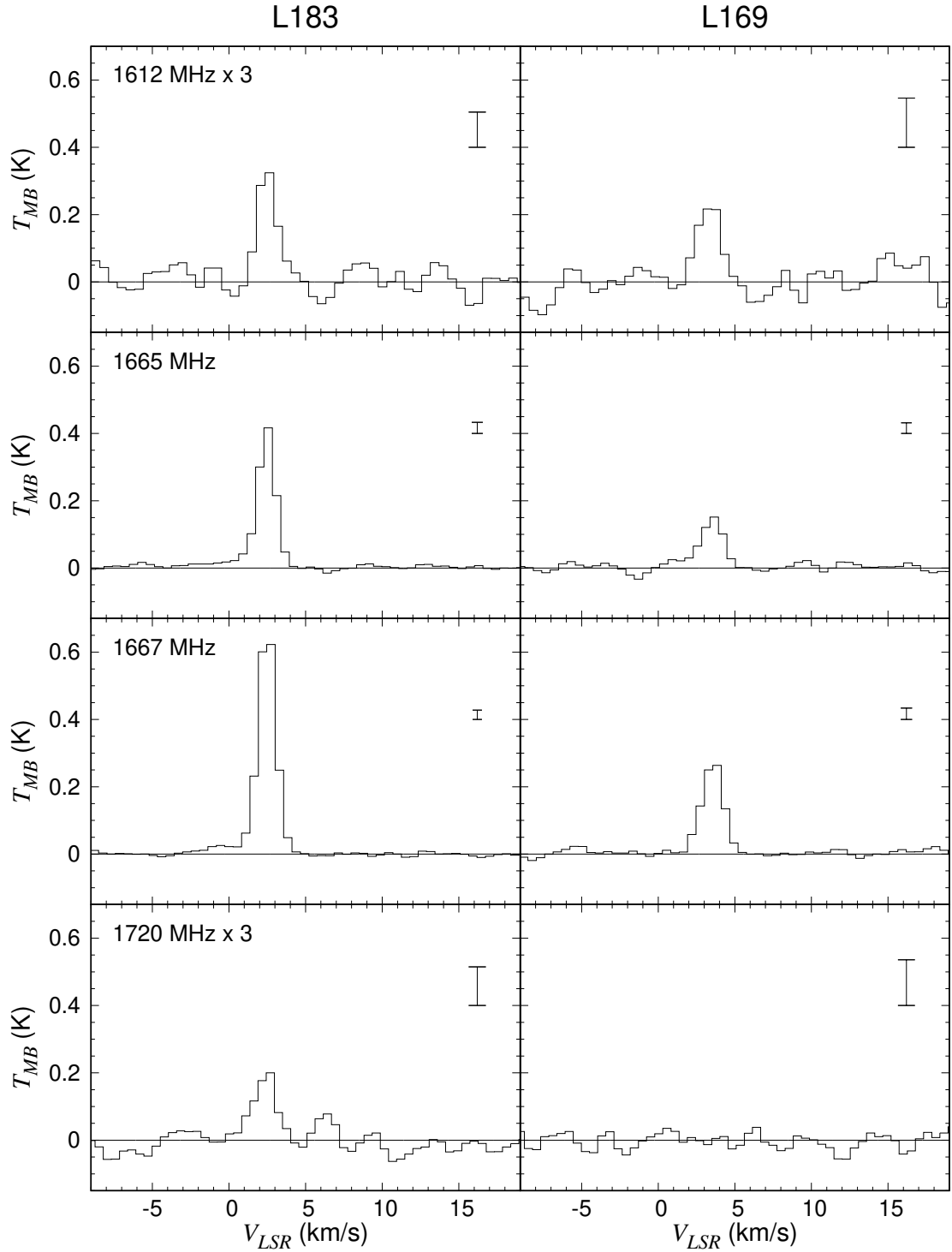}
  \caption{
    Observed spectra of the OH 18 cm transition toward L183 (left) and L169 (right).
    Three times the rms noise of each spectrum is represented in the top right of each panel.
    \label{fig:sL183}
  }
\end{figure}

    A similar hfs intensity anomaly is also observed near the cold starless core TMC-1(CP) in HCL2 (Figure $\ref{fig:HCL2}$).
    Figures $\ref{fig:sTMC1CP_0_24}$--$\ref{fig:sTMC1CP_0_8}$ show the observed spectra of the OH 18 cm transition toward three positions located 24$'$, 16$'$, and 8$'$ north from TMC-1(CP), respectively, as indicated in Figure $\ref{fig:HCL2}$.
    The 1612 MHz line is observed in faint emission or slight absorption for the red-shifted component ($V_{LSR}$=6--10 km s$^{-1}$), suggesting that this velocity component traces warm gas.
    On the other hand, the 1720 MHz line is much fainter than the 1612 MHz line for the blue-shifted component ($V_{LSR}$ $\sim$ 5 km s$^{-1}$),
    although these two lines should have the same intensity in LTE ($I_{1612}:I_{1665}:I_{1667}:I_{1720}=1:5:9:1$).
    \citet{Harju2000} also reported that the 1720 MHz line is fainter than the 1612 MHz line toward the north part of TMC-1 ridge.
    Again, this anomalous feature could not be reproduced with our statistical equilibrium calculation without considering FIR pumping.
    Figure $\ref{fig:sL183}$ (left) shows the observed OH spectra toward L183 (Figure $\ref{fig:L183}$), which we already reported in our previous paper \citep{Ebisawa2015}.
    The 1720 MHz line is slightly weaker than the 1612 MHz line toward L183.
    This feature is more evident in the dense core L169, which is adjacent to L183 (Figure $\ref{fig:L183}$).
    As shown in Figure $\ref{fig:sL183}$ (right), the 1720 MHz line is totally absent toward L169, whereas the 1612 MHz line shows bright emission.
    An intensity ratio of the 1612 MHz line to the 1667 MHz line
    is about 1/3, which is three times higher than the expected value of 1/9 in LTE.
    The line parameters are summarized in Table $\ref{table:linepara}$.

    As described above, the 1720 MHz line absorption (or its faint emission) and the enhanced 1612 MHz line emission are observed toward several sources.
    Thus, revealing the origin of this anomaly is important for a full understanding of the excitation mechanism of OH.
    For this purpose, here we investigate the effect of far-infrared pumping of OH.
\section{EFFECT OF FIR PUMPING} \label{sec:Analysis}
\begin{figure}[htbp] 
  \centering
  \epsscale{1}
  \plotone{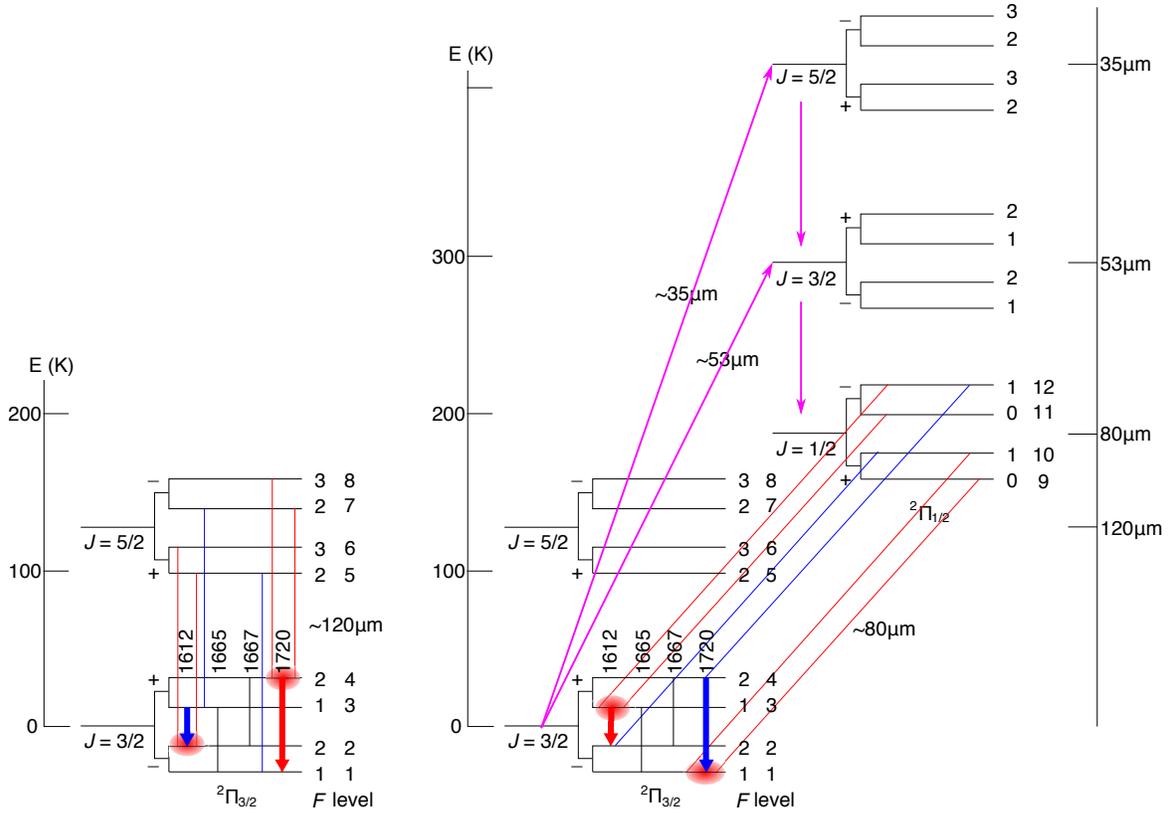}
  \caption{%
    Schematic illustrations of the intensity anomaly of the OH 18 cm transition.
    (left) The transition from the first rotationally excited state to the ground state levels causes overpopulation in the $F=2$ levels in the ground state,
    and the 1612 MHz line becomes weaker than expected under LTE.
    The 1720 MHz line becomes brighter at the same time.
    (right) The transition from the second rotationally excited state to the ground state levels works in the opposite way,
    and the $F=1$ levels of the ground state become overpopulated.
    The 1720 MHz line and the 1612 MHz line become weaker and brighter, respectively, in this case.
    The populations in the second rotationally excited state levels can be enhanced
    by the FIR pumping from the ground state levels
    to the $^2\Pi_{1/2}$ $J=3/2$ and $^2\Pi_{1/2}$ $J=5/2$ state levels
    at wavelengths of 53 and 35 $\mu$m, respectively.
    The OH molecules are de-excited to the second rotationally excited state
    by the subsequent spontaneous emission, as represented by magenta arrows.
    \label{fig:OH_energy_FIR}}
\end{figure}

    The hyperfine intensity anomaly is caused by the excitation and de-excitation between the rotationally excited states and the ground state,
    as schematically illustrated in Figure $\ref{fig:OH_energy_FIR}$.
    Emission from the first rotationally excited state ($J=5/2$ of $^2\Pi_{3/2}$) to the ground state ($J=3/2$ of $^2\Pi_{3/2}$)
    produces an overpopulation in the $F=2$ levels in the ground state due to the selection rule, as described in Figure $\ref{fig:OH_energy_FIR}$ (left).
    This anomalous population decreases the excitation temperature of the 1612 MHz line and increases that of the 1720 MHz line.
    This leads to the 1612 MHz line absorption against the CMB.
    On the other hand, emission from the second rotationally excited state ($J=1/2$ of $^2\Pi_{1/2}$) to the ground state levels works in the opposite way (Figure $\ref{fig:OH_energy_FIR}$, right).
    In this case, $F=1$ levels in the ground state are overpopulated, increasing the excitation temperature of the 1612 MHz line and decreasing that of the 1720 MHz line.
    Thus the 1720 MHz line can appear in absorption if this trend is strong.
    In different parts of the ISM, both processes may occur \added{{with different contributions}},
    but the collisional excitation with H$_2$ to the first rotationally excited state levels is more efficient than that to the second rotationally excited state.
    The first effect is therefore dominant, producing the absorption of the 1612 MHz line.

    Nevertheless, the second effect can be dominant,
    if FIR pumping to the $^2\Pi_{1/2}$ $J=3/2$ and $^2\Pi_{1/2}$ $J=5/2$ states
    at wavelengths of 53 $\mu$m and 35 $\mu$m, respectively, is efficient.
    Most of the OH molecules excited to these states radiatively decay to the
    $^2\Pi_{1/2}$ $J=1/2$ state by subsequent rotational transitions
    within the $^2\Pi_{1/2}$ ladder,
    as represented by magenta arrows in Figure $\ref{fig:OH_energy_FIR}$.
    This is because the spontaneous emission rates of the transitions
    within the $^2\Pi_{1/2}$ ladder are typically one order of magnitude \replaced{larger}{{higher}}
    than those
    of the inter-ladder transitions between the $^2\Pi_{1/2}$ and $^2\Pi_{3/2}$ states.
    Then, rotational transitions from the $^2\Pi_{1/2}$ $J=1/2$ state to the
    rotational ground state might produce the 1720 MHz line absorption, as described before.
    It should be noted that the direct FIR pumping from the $^2\Pi_{3/2}$ $J=3/2$ ground state
    to the $^2\Pi_{1/2}$ $J=1/2$ state at 80 $\mu$m does not contribute to the population anomaly,
    because the subsequent radiative de-excitation back to the ground state occurs under the same selection rules
    as the FIR pumping.
    These mechanisms are indeed confirmed by our simulations,
    as described in Section $\ref{sec:analysis_Tk}$.

    The effect of FIR pumping on the excitation of OH was studied by \citet{Elitzur1976a}.
    These authors showed that the 1612 MHz line becomes stronger and will display maser emission in extreme cases,
    particularly in the dusty envelopes of stars, so-called OH-IR stars.
    They reported that the 1612 MHz maser can be explained
    by the FIR pumping to the $^2\Pi_{1/2}$ $J=5/2$ state by 35 $\mu$m photons emitted
    from warm dust grains ($T_d$ = 100--200 K).
    While no bright infrared source is associated with our observed sources,
    the FIR radiation from dust grains in surrounding warm clouds
    would contribute to the pumping to the
    $^2\Pi_{1/2}$ \replaced{$J=3/2$ and $^2\Pi_{1/2}$ $J=5/2$ states}{{ladder}}.
    Therefore, we incorporate the effect of the FIR radiation into our statistical equilibrium calculations to examine whether we can explain the 1720 MHz line absorption
    \replaced{and make inferences on}{{, and to constrain}} the physical conditions in the OH bearing clouds.\deleted{.}

    In addition, we consider line overlaps by following the formulation of \citet{Guilloteau1981}.
    They incorporated line overlaps of the far-infrared OH rotational transitions in their statistical equilibrium calculations,
    assuming a uniform condition in a spherical cloud,
    and reported that the pumping effect due to the overlaps well explains
    the intensities of strong OH masers found in the dense envelopes of compact H$_{\rm II}$ regions.

    \subsection{Statistical equilibrium calculation considering the FIR pumping effect}
        \subsubsection{Formulation}
            We incorporate the pumping effect by the FIR radiation from dust grains into our statistical equilibrium calculation \citep{Ebisawa2015} in a similar way to that reported by \citet{Elitzur1976a}.
            We also consider the effect of line overlaps approximately, as described by \citet{Guilloteau1981}.
            The statistical equilibrium takes the form
            \begin{eqnarray}
              \frac{d(g_in_i)}{dt} &=&
                \sum_{j>i} \left\{
                  g_j {\beta}_{ji} A_{ji} \left[
                    n_j
                    - \left(
                      R_{ji} + R^d_{ji} \right)
                    \left(
                      n_i - n_j \right)
                    \right]\right.\nonumber\\
                  &+& \left. \sum_{k=(u_k, l_k)} \left(
                      g_j A_{ji} P_{(ji), (u_kl_k)} n_j - g_{u_k} A_{u_k l_k} P_{(u_kl_k), (ji)}n_{u_k}
                      \right)\right.\nonumber\\
                  &+& \left. g_j C_{ji} \left[
                    n_j - n_i \exp \left(
                      - \frac{h{\nu}_{ji}}{k_bT_k} \right)
                    \right]
                  \right\}\nonumber\\
                &-& \sum_{i>j} \left\{
                  g_i {\beta}_{ij} A_{ij} \left[
                    n_i
                    - \left(
                      R_{ij} + R^d_{ij} \right)
                    \left(
                      n_j - n_i \right)
                    \right]\right.\nonumber\\
                  &+& \left.\sum_{k=(u_k, l_k)} \left(
                      g_i A_{ij} P_{(ij), (u_kl_k)} n_i - g_{u_k} A_{u_k l_k} P_{(u_kl_k), (ij)}n_{u_k}
                      \right)\right.\nonumber\\
                  &+& \left. g_i C_{ij} \left[
                    n_i - n_j \exp \left(
                      -\frac{h{\nu}_{ij}}{k_bT_k} \right)
                    \right]
                  \right\}\nonumber\\
                &=& 0,
                \label{eq:equilibrium}
            \end{eqnarray}
            where $T_k$, $k_b$, and $h$ are the gas kinetic temperature,
            the Boltzmann constant, and the Planck constant, respectively.
            $n_i$ and $g_i$ represent the population and the degeneracy of the energy level $i$ of OH, respectively.
            $A_{ij}$, $C_{ij}$, and ${\nu}_{ij}$ are the Einstein A coefficient,
            the collisional rate, and the frequency of the transition, respectively, between the levels $i$ and $j$;
            $k=(u_k, l_k)$ in the second and fifth terms of Equation ($\ref{eq:equilibrium}$)
            denotes summation over all transitions of OH,
            whereas $j$ in the first and fourth terms of Equation ($\ref{eq:equilibrium}$)
            represents summation over all levels.
            $R_{ij}$ stands for the photon occupation number of the cosmic microwave background (CMB) radiation at the frequency of the transition between the levels $i$ and $j$ as
            \begin{eqnarray}
              R_{ij} &=& \frac{1}{\exp\left(h{\nu}_{ij}/k_bT_{CMB}\right)-1},
            \end{eqnarray}
            where $T_{CMB}$ equals 2.73 K.

            $R^d_{ij}$ is the photon occupation number of the dust FIR radiation at the frequency of the transition between the levels $i$ and $j$.
            We examine two models for the derivation of $R^d_{ij}$.
            First, we assume that the dust grains emit gray-body radiation of dust temperature $T_d$ and spectral index ${\beta}_d$.
            In this case, $R^d_{ij}$ can be written as
            \begin{eqnarray}
              R^d_{ij} &=& \left(\frac{\nu}{{\nu}_{160}}\right)^{{\beta}_d}\frac{\eta}{\exp\left(h{\nu}_{ij}/k_bT_{d}\right)-1},
              \label{eq:graybody}
            \end{eqnarray}
            where $\eta$ is a dilution factor of the dust thermal
            radiation\replaced{, which}{{. It}} is derived from the estimated intensity of the FIR radiation at 160 $\mu$m ($I_{160}$) divided by the intensity of the blackbody radiation (times ${\nu}^{{\beta}_d}$) at this wavelength.
            We perform the calculation for ${\beta}_d$ values from 0 to 3.
            Although a higher value of ${\beta}_d$ gives a better fit to the observed spectra,
            there is no significant dependence on ${\beta}_d$ as long as it is higher than 2.
            Therefore, ${\beta}_d$ is assumed to be 2, which has been reported as a typical value for the diffuse ISM \citep[e.g.][]{Boulanger1996}.

            Second, we use DustEM to model the dust emission \citep{Compiegene2011}
            ( https://www.ias.u-psud.fr/DUSTEM/index.html) in order to derive $R_{ij}^d$.
            We employ the DustEM version calculated by \citet{Jones2013}.
            As shown later, the FIR emission predicted by the DustEM model better reproduces the 1720 MHz line absorption of OH
            than the gray-body approximation, although it depends on the dust temperature.

            ${\beta}_{ij}$ in the first and fourth terms of Equation ($\ref{eq:equilibrium}$) denotes the photon escape probability of the transition between the levels $i$ and $j$.
            $P_{(ij),(kl)}$ is the probability that a photon emitted in the transition
            between the levels $i$ and $j$ is absorbed by that between the levels $k$ and $l$,
            which represents the effect of the line overlaps.
            We follow the formulation of ${\beta}_{ij}$ and $P_{(ij), (kl)}$ introduced by \citet{Guilloteau1981}.

        \subsubsection{FIR model}\label{sec:FIRmodel}
\begin{figure}[htbp] 
    \centering
    \epsscale{1.0}
    \plotone{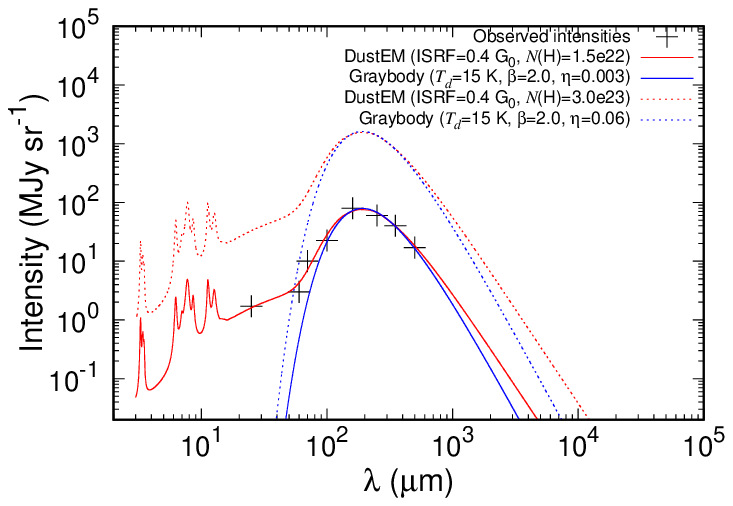}
    \caption{
        (Cross) Intensities of the mid- and far-infrared radiation observed toward TMC-1FN.
        The intensities are taken from
        Spitzer 70 and 160 $\mu$m maps by \citet{Flagey2009},
        IRAS 25, 60 and 100 $\mu$m maps by \citet{Flagey2009},
        and Herschel 250, 350 and 500 $\mu$m archival data
        with observation IDs of 1342202252 and 1342202253.
        (Solid red line) The best-fitting DustEM \citep{Jones2013} profile assuming the ISRF of 0.4 $G_0$,
        which gives the best-fitting result among the ISRF values from 0.1 $G_0$ to 1.0 $G_0$ at a 0.1 $G_0$ interval.
        (Solid blue line) The gray-body profile that best explains the above DustEM model,
        assuming $\beta$ of 2. The dust temperature ($T_d$) is determined to be 15 K.
        The observed intensities in 25, 60, and 70 $\mu$m are stronger than those estimated by the gray-body,
        whereas they are well reproduced by the DustEM model.
        The best-fitting DustEM and gray-body profiles are scaled by a factor of $\sim$20,
        \replaced{which}{{and}} are employed in our statistical equilibrium calculations for the fiducial models,
        as shown by red and blue dotted lines, respectively.
        In Figures $\ref{fig:LVG_FIR}$ and $\ref{fig:Ndust_nH2}$,
        this scaling is represented by the flux at 160 $\mu$m.
        \label{fig:SED_fit}}
\end{figure}

            The FIR pumping effect is incorporated by using the DustEM model by \citet{Jones2013} and the gray-body profile.
            In the DustEM model, the interstellar radiation field (ISRF) strength
            and the H column density are given as the parameters.
            We employ the ISRF strength of 0.4 G$_0$,
            because it best explains the shape of SED observed toward TMC-1FN \citep{Flagey2009}
            among the ISRF values from 0.1 $G_0$ to 1.0 $G_0$ at a 0.1 $G_0$ interval
            (red solid line in Figure $\ref{fig:SED_fit}$).
            Here $G_0$ stands for a scaling factor of the UV field with respect to
            the flux in the Habing field \citep{Habing1968}.
            The H column density is also determined to be 1.5 $\times$ 10$^{22}$ cm$^{-2}$.
            However, we use the H column density of 3 $\times$ 10$^{23}$ cm$^{-2}$
            in the fiducial DustEM model for the statistical equilibrium calculations
            by scaling the FIR intensity by a factor of 20 (red dotted line in Figure $\ref{fig:SED_fit}$)
            so that the 1720 MHz absorption is reproduced.
            Apparently, this H column density is too high for the line-of-sight value by one order of magnitude.
            We will discuss this point later in Section $\ref{sec:TMC-1FN}$.
            In the gray-body model, the dust temperature ($T_d$) and the dilution factor ($\eta$)
            are estimated to be 15 K and 0.003, respectively,
            by fitting the above DustEM model for the H column density of 1.5 $\times$ 10$^{22}$ cm$^{-2}$
            (blue solid line in Figure $\ref{fig:SED_fit}$).
            As in the case for the fiducial DustEM model, this gray-body profile is multiplied by a factor of 20
            (blue dotted line), which is employed in our statistical equilibrium calculations.

            It should be noted that the observed intensities at 25, 60 and 70 $\mu$m
            are slightly higher than those expected by the gray-body.
            Although the discrepancy is marginal in Figure $\ref{fig:SED_fit}$,
            it is more evident in the mean SED of Taurus molecular cloud complex
            averaged over a 44 deg$^2$ area including TMC-1FN \citep{Flagey2009}.
            As shown in the solid red line in Figure $\ref{fig:SED_fit}$,
            the DustEM model well reproduces the excess feature at shorter wavelengths
            compared to the gray-body.
            The FIR intensities at 35, 53 and 120 $\mu$m are predicted to be
            2.2, 3.3 and 48 MJy sr$^{-1}$, respectively,
            by the DustEM model for the solid red line of Figure $\ref{fig:SED_fit}$,
            while they are predicted to be
            8.5 $\times$ 10$^{-5}$, 0.1 and 42 MJy sr$^{-1}$,
            respectively, with the gray-body for the solid blue line of Figure $\ref{fig:SED_fit}$.
            Namely, the FIR intensities calculated by the DustEM model
            at 35 and 53 $\mu$m relative to the 120 $\mu$m intensity
            are 2 $\times$ 10$^4$ and 29 times larger, respectively, compared to those
            calculated by the gray-body model.
            Higher intensities in the 35 $\mu$m and 53 $\mu$m contribute to
            more efficient pumpings to the $^2\Pi_{1/2}$ $J=5/2$ and $^2\Pi_{1/2}$ $J=3/2$ states,
            respectively, which leads to the 1720 MHz line absorption,
            as described in Section $\ref{sec:Analysis}$.
        \subsubsection{Condition for the 1720 MHz line absorption}\label{sec:analysis_color}
\begin{figure}[htbp] 
  \centering
  \epsscale{1.2}
  \plotone{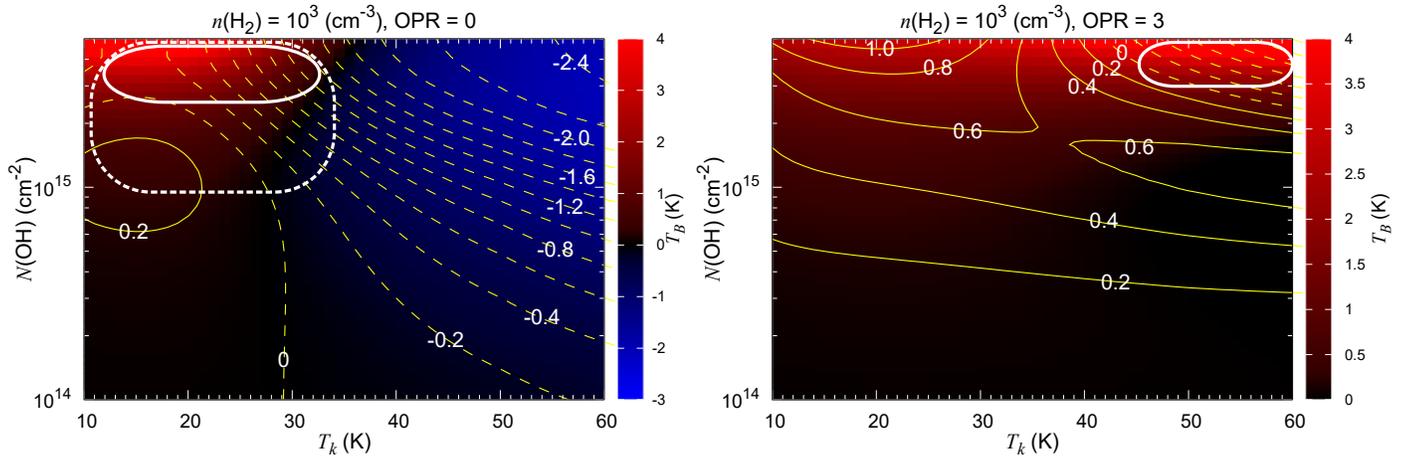}
  \caption{%
    Intensities of the 1612 MHz line (color) and the 1720 MHz line (yellow contours)
    on the $N$(OH)--$T_k$ plane derived from our statistical equilibrium calculation,
    considering the effect of the FIR radiation.
    (Left) The H$_2$ ortho-to-para ratios is assumed to be 0.
    Contour levels are shown for the 1720 MHz line intensity from -2.4 to 0.2 K with 0.2 K interval.
    Dashed contours correspond to negative brightness temperatures of the 1720 MHz line.
    The 1720 MHz line absorption and the bright emission of the 1612 MHz line can be reproduced
    for $N$(OH) $>$ 3 $\times$ 10$^{15}$ cm$^{-2}$ and $T_k <$ 30 K (white solid circled area).
    The case when the 1720 MHz line is brighter than the 1612 MHz line can be reproduced
    for $N$(OH) $>$ 10$^{15}$ cm$^{-2}$ and $T_k <$ 30 K (white dotted circled area).
    (right) The H$_2$ ortho-to-para ratios is assumed to be 3.
    Contour levels are shown for the 1720 MHz line intensity from -0.8 K to 1.0 K with 0.2 K interval.
    The 1720 MHz line absorption and the bright emission of the 1612 MHz line can be reproduced
    for $N$(OH) $>$ 2 $\times$ 10$^{15}$ cm$^{-2}$ and $T_k >$ 40 K (white circled area).
    \label{fig:LVGcolormap}}
\end{figure}

            Figure $\ref{fig:LVGcolormap}$ shows the intensities of the 1612 MHz line (color)
            and the 1720 MHz line (contours) derived from our statistical equilibrium calculation
            in the $N$(OH) -- $T_k$ plane.
            The FIR pumping effect is treated with the DustEM model by \citet{Jones2013} assuming an
            ISRF strength of
            0.4 $G_0$ and an H column density of 3 $\times$ 10$^{23}$ cm$^{-2}$
            (red dotted line of Figure $\ref{fig:SED_fit}$).
            The H$_2$ density is assumed to be 10$^3$ cm$^{-3}$ as a representative value.
            The H$_2$ ortho-to-para ratio is assumed to be 0 and 3 in the left and right panels, respectively.
            Dependencies on these parameters are described later.
            The linewidth is assumed to be a typical observed value of 1.0 km s$^{-1}$.

            Dashed contours in Figure $\ref{fig:LVGcolormap}$ represent absorption of the 1720 MHz line,
            whereas the blue colored region corresponds to absorption of the 1612 MHz line.
            As shown in Figure $\ref{fig:LVGcolormap}$ (left), the case for which the emission in the 1612 MHz line
            is stronger than that in the 1720 MHz line can be reproduced for parameters contained in the area approximately enclosed by the dotted lines.
            Namely, an OH column density higher than $\sim$10$^{15}$ cm$^{-2}$ and a gas kinetic temperature lower than $\sim$30 K are necessary if the H$_2$ ortho-to-para ratio is 0.
            A higher OH column density ($N$(OH) $>$ 3 $\times$ 10$^{15}$ cm$^{-2}$) is required to reproduce the 1720 MHz line absorption,
            as represented by the white solid lines.
            The reason why such a high column density is needed for the 1720 MHz line absorption
            is that the transitions from the second rotationally excited state to the ground state
            must be optically thick in order to produce the anomalous population
            by the mechanism described in Figure $\ref{fig:OH_energy_FIR}$.
            Assuming an H$_2$ density of 10$^{3}$ cm$^{-3}$ and an OH column density of 3 $\times$ 10$^{15}$ cm$^{-2}$,
            the fractional abundance of OH is estimated to be 10$^{-6}$,
            which is an order of magnitude higher than the typical value of the fractional abundance
            in diffuse clouds \citep{Wiesemeyer2012}.
            However, the fractional abundance of OH is enhanced in the shocked gas,
            and it can reach values as high as $\sim$ 5 $\times$ 10$^{-5}$ according to \citet{Draine1986}.
            Hence the 1720 MHz line could appear in absorption in such an OH-rich gas
            that experienced the shock.
            It should be noted that the case in which the 1720 MHz line is weaker than the 1612 MHz line can be reproduced,
            even with a typical OH fractional abundance of 10$^{-7}$.

\begin{figure}[htbp] 
  \centering
  \epsscale{1.0}
  \plotone{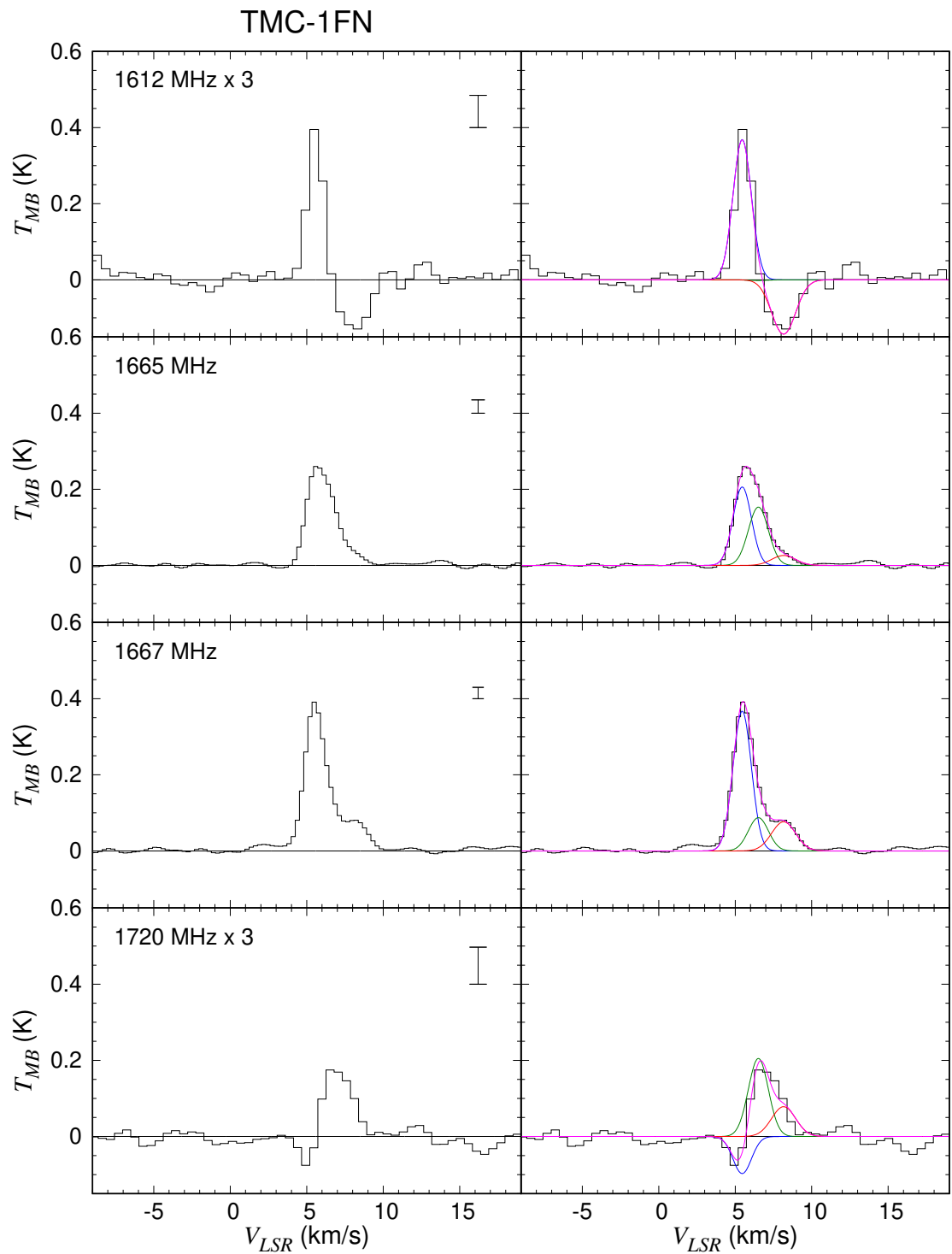}
  \caption{
    (Left) Observed spectra of the OH 18 cm transition toward TMC-1FN.
    Three times the rms noise of each spectrum is represented in the top right.
    (Right) The best-fit Gaussian profiles obtained
    by assuming three velocity components shown by blue, green, and red lines,
    whose central velocities are 5.5, 6.5, and 8.2 km s$^{-1}$, respectively.
    Magenta lines \replaced{shows}{{show}} the sum of these three velocity components.
    \label{fig:sTMC1FN}
  }
\end{figure}

            Intensities of the 1612 MHz and the 1720 MHz lines are determined to be
            $\sim$1--4 K and $-$(0.1--0.4) K within the solid circle in Figure $\ref{fig:LVGcolormap}$ (left),
            which are about one order of magnitude stronger
            than those observed toward TMC-1FN (Figure $\ref{fig:sTMC1FN}$).
            It can be explained by the beam-dilution effect with a beam filling factor of $\sim$0.1,
            a plausible value since the 1720 MHz line absorption traces
            a relatively cold and dense structure, as shown in the following subsections.

            For an H$_2$ ortho-to-para ratio of 3 (Figure $\ref{fig:LVGcolormap}$, right),
            the 1720 MHz line absorption is reproduced
            for an OH column density of 10$^{15}$ cm$^{-2}$ or higher
            and a gas kinetic temperature of 40 K or higher
            (white circle in Figure $\ref{fig:LVGcolormap}$, right).
            This absorption feature is the result of line overlaps,
            as shown later in Section $\ref{sec:analysis_Tk}$.
            In particular, line overlaps between the transitions
            from the first rotationally excited state to the
            \replaced{rotational ground}{{ground rotational}} state
            play an important role,
            where the overlaps between the two transitions from the first rotationally excited state
            to the $F=2$ levels in the \replaced{rotational ground}{{ground rotational}} state
            ---that is, transitions marked (6, 2) and (5, 2), and (8, 4) and (7, 4) in Table $\ref{table:overlap}$---
            are significant.
            Here, a typical observed linewidth of 1.0 km s$^{-1}$ is considered.
            In this situation, the transitions to the $F=2$ levels can be readily saturated,
            while those to the $F=1$ levels are not.
            Hence, the $F=1$ levels in the \replaced{rotational ground}{{ground rotational}} state are relatively overpopulated,
            which makes the 1720 MHz line weaker.
            The 1720 MHz line can appear in absorption, if this trend is strong.

            It should be noted that the line overlaps among the transitions
            from the second rotationally excited state to the ground state
            do not seriously contribute to the hfs anomaly.
            For instance,
            the overlap between transitions (10, 1) and (9, 1),
            which we hereafter term the (10, 1)--(9, 1) overlap,
            produces the overpopulation in the level 2 ($F=2$) in the \replaced{rotational ground}{{ground rotational}} state.
            It makes the 1720 MHz line brighter.
            However, this overpopulation is less effective
            than the overpopulation in the level 1 ($F=1$)
            produced by the contribution of the first rotationally excited state, i.e., the (6, 2)--(5, 2) overlap.
            This is because the former overpopulation
            is suppressed by the (10, 2)--(10, 1) and (10, 2)--(9, 1) overlaps.
            Such a phenomenon also occurs in the other overlaps,
            and eventually the contribution of the line overlaps among the transitions from the
            second rotationally excited state to the anomaly is limited.


            The collisional excitations to the first rotationally excited state
            occur more efficiently with a higher gas kinetic temperature and a higher H$_2$ ortho-to-para ratio.
            In addition, transitions from the first rotational excited states to the \replaced{rotational ground}{{ground rotational}} state must be optically thick
            in order to produce the anomalous population by the above mechanism.
            Hence, a high OH column density ($>$ 10$^{15}$ cm$^{-2}$)
            and a high gas kinetic temperature ($>$ 40 K)
            are required to reproduce the absorption feature of the 1720 MHz line
            (Figure $\ref{fig:LVGcolormap}$, right).
            However, the 1612 MHz line is much brighter than the 1665 MHz line
            when the 1720 MHz line appears in absorption for an H$_2$ ortho-to-para ratio of 3,
            as shown in Figure $\ref{fig:SEC_N15_OPR3_Tk}$ of Appendix A,
            which is inconsistent with the observation.
            Therefore, we assume an H$_2$ ortho-to-para ratio of 0,
            an OH column density of 3 $\times$ 10$^{15}$ cm$^{-2}$,
            and a gas kinetic temperature of 20 K (Figure $\ref{fig:LVGcolormap}$, left)
            for our fiducial model in the following sections.
            Results for an H$_2$ ortho-to-para ratio of 3
            and an OH column density of 3 $\times$ 10$^{14}$ cm$^{-2}$
            are presented in Appendix A, for reference.

        \subsubsection{Dependence on FIR intensity and mechanism of the 1720 MHz absorption} \label{sec:analysis_Tk}
            Figure $\ref{fig:SEC_N15_Tk}$ shows the calculated intensities of the OH 18 cm transition
            as a function of the gas kinetic temperature
            for the three cases of the FIR intensity.
            As noted above, the OH column density and the H$_2$ ortho-to-para ratio are assumed
            to be 3$\times$10$^{15}$ cm$^{-2}$ and 0, respectively.
            The effect of the line overlaps is not considered in the left panels (a)--(c),
            whereas it is considered in the right panels (d)--(f).
            No FIR radiation is assumed to be present in the calculations illustrated in the top panels ((a) and (d)),
            whereas the gray-body approximation with a dust temperature of 15 K
            applies for the middle panels ((b) and (e)).
            The FIR radiation calculated by the DustEM \citep{Jones2013}
            is employed in the bottom panels ((c) and (f)).
            An ISRF strength of
            0.4 $G_0$ and an H column density of 3 $\times$ 10$^{23}$ cm$^{-2}$
            are assumed for the fiducial DustEM model,
            same as assumed for Figure $\ref{fig:LVGcolormap}$.

            As shown in Figure $\ref{fig:SEC_N15_Tk}$ (f),
            the 1720 MHz line absorption and the 1612 MHz line emission are reproduced
            by the fiducial DustEM model considering the effect of the line overlaps
            for gas kinetic temperatures lower than 30 K,
            which is consistent with the results in Section $\ref{sec:analysis_color}$
            (see the orange area of Figure $\ref{fig:SEC_N15_Tk}$ (f)).
            The effect of the line overlaps produces a weaker 1720 MHz line,
            which is clearly seen by comparing
            the left ((a), (b), (c)) and right ((d), (e), (f)) panels of Figure $\ref{fig:SEC_N15_Tk}$.
            All the four hfs lines of OH appear in absorption
            for gas kinetic temperatures higher than about 30 K.
            In contrast to the fiducial DustEM model, the absorption only
            in the 1720 MHz line is not reproduced
            by the model employing the gray-body approximation ((b) and (e)),
            although the 1720 MHz line becomes weaker than for the calculations
            assuming no FIR radiation ((a) and (d)).
            This is because the excitation to the
            $^2\Pi_{1/2}$ $J=3/2$ and $^2\Pi_{1/2}$ $J=5/2$ states
            via FIR radiation is more efficient in the DustEM model
            than in the gray-body model,
            due to an excess of the FIR radiation at shorter wavelengths in the DustEM model,
            as described in Section $\ref{sec:FIRmodel}$ (Figure $\ref{fig:SED_fit}$).

            In order to confirm this,
            we conduct the additional statistical equilibrium calculations
            assuming the same conditions of Figure $\ref{fig:SEC_N15_Tk}$ (f),
            but also applying an artificial wavelength threshold ($\lambda_{th}$) for the FIR radiation,
            where no FIR radiation is assumed for wavelengths shorter than $\lambda_{th}$.
            Figures $\ref{fig:SEC_N15_Tk_fcut}$ (a)--(d) show the results
            with $\lambda_{th}$ values of 60, 50, 30 and 0 $\mu$m, respectively.
            The panel (d) of Figure $\ref{fig:SEC_N15_Tk_fcut}$ ($\lambda_{th}$=0 $\mu$m)
            is the same as the panel (f) of Figure $\ref{fig:SEC_N15_Tk}$.
            In the case of the $\lambda_{th}$ of 60 $\mu$m (a),
            neither the 53 $\mu$m nor the 35 $\mu$m pumping effect is included,
            and hence, the 1720 MHz line absorption is not reproduced.
            On the other hand, the 1720 MHz line becomes very faint with
            a gas kinetic temperature lower than 30 K in the panel (b) ($\lambda_{th}$=50 $\mu$m).
            This is caused by the effect of 53 $\mu$m pumping, although the absorption feature is not reproduced.
            The intensity of the 1720 MHz line is further decreased by the contribution
            from the 35 $\mu$m pumping effect in the panel (c) ($\lambda_{th}$=30 $\mu$m),
            and the absorption feature is reproduced.
            The calculated OH hfs intensities in the panel (c) are almost comparable
            to those in the panel (d) ($\lambda_{th}$=0 $\mu$m).
            Therefore, the 1720 MHz line absorption can be explained
            by the FIR pumping effect in 53 $\mu$m and 35 $\mu$m,
            according to Figure $\ref{fig:SEC_N15_Tk_fcut}$.

\begin{figure}[htbp] 
  \centering
  \epsscale{1.2}
  \plotone{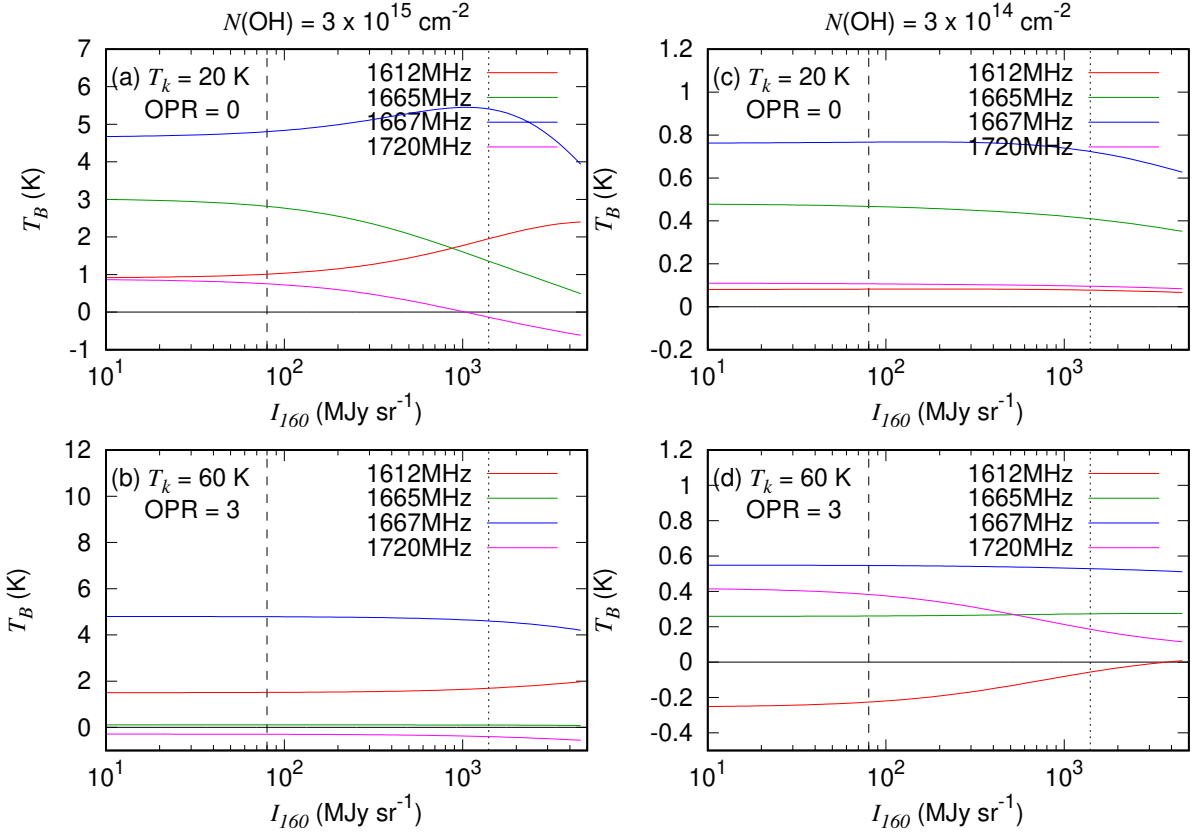}
  \caption{
    Intensities of the hfs lines of the OH 18 cm transition
    as a function of the intensity of the dust thermal emission
    at 160 $\mu$m ($I_{160}$) derived from our statistical equilibrium calculation.
    $I_{160}$ is used as a parameter to scale the FIR intensity profile of the DustEM model (Figure $\ref{fig:SED_fit}$).
    The dotted lines show $I_{160}$ for the fiducial DustEM model,
    while the dashed lines show $I_{160}$ observed toward TMC-1FN.
    The H$_2$ density is assumed to be 10$^3$ cm$^{-3}$.
    The column density of OH is assumed to be 3 $\times$ 10$^{15}$ cm$^{-2}$ (left) or 3 $\times$ 10$^{14}$ cm$^{-2}$ (right).
    The gas kinetic temperature and H$_2$ ortho-to-para ratio are assumed to be 20 K and 0, respectively, in the top panels,
    whereas they are assumed to be 60 K and 3, respectively, in the bottom panels.
    \label{fig:LVG_FIR}}
\end{figure}

            Figure $\ref{fig:LVG_FIR}$ summarizes the above results
            as a function of the intensity of the FIR radiation at 160 $\mu$m ($I_{160}$)
            for four combinations of an OH column density and a gas kinetic temperature.
            We employ the $I_{160}$ as a representative value to scale the FIR intensity,
            since the observed SED is the strongest at this wavelength.
            It should be noted that the $I_{160}$ is estimated to be 1.4 $\times$ 10$^3$ MJy sr$^{-1}$
            with the fiducial DustEM model assuming an ISRF of 0.4 G$_0$ and an H column density of
            3 $\times$ 10$^{23}$ cm$^{-2}$ (red dotted line in Figure $\ref{fig:SED_fit}$).
            The H$_2$ ortho-to-para ratio is set to be 0 for the gas kinetic temperature of 20 K,
            while it is set to be 3 for the temperature of 60 K.
            The H$_2$ density is assumed to be 10$^3$ cm$^{-3}$.
            The DustEM model is applied for the FIR radiation, and the effect of the overlaps is considered.
            As shown in Figure $\ref{fig:LVG_FIR}$, the effect of the FIR radiation is pronounced for two parameter combinations:
            high OH column density (3 $\times$ 10$^{15}$ cm$^{-2}$) together with low gas kinetic temperature (20 K) (panel (a)),
            and
            low OH column density (3 $\times$ 10$^{14}$ cm$^{-2}$) and high gas kinetic temperature (60 K) (panel (d)).
            The 1720 MHz line becomes weaker and the 1612 MHz line becomes stronger
            for increasing FIR radiation intensity in both cases.
            In the former case, the 1720 MHz line appears in absorption
            for an FIR intensity larger than 10$^3$ MJy sr$^{-1}$ (Figure $\ref{fig:LVG_FIR}$ (a)).
            Note that the 1720 MHz line absorption is also reproduced for
            $T_k$=60 K and $N$(OH)=3$\times$10$^{15}$ cm$^{-2}$ (panel (b)).
            This absorption is considered to be caused by the effect of the line overlaps
            through the mechanism explained in Section $\ref{sec:analysis_color}$,
            as the intensities of the OH 18 cm transition
            are insensitive to the FIR intensity (Figure $\ref{fig:LVG_FIR}$ (b)).

            In our previous paper \citep{Ebisawa2015}, we reported that the OH 18 cm transition can be used as a
            thermometer for diffuse ($N$(OH)$\sim$10$^{14}$ cm$^{-2}$)
            and warm ($T_k$$\sim$50 K) molecular gas.
            As shown in Figure $\ref{fig:LVG_FIR}$ (d), the effect of the FIR radiation is negligible
            for an 160 $\mu$m FIR intensity lower than about 10$^2$ MJy sr$^{-1}$
            under diffuse warm gas conditions ($T_k$=60 K, $N$(OH)=3$\times$10$^{14}$ cm$^{-2}$).
            On the other hand, the absolute intensity of the 1612 MHz line absorption is decreased by $\sim$50\%
            for $I_{160}=500$ MJy sr$^{-1}$.
            This decrease of the 1612 MHz line intensity can cause a systematic error of $\sim$10 K in the gas kinetic temperature derived from our statistical equilibrium calculations, assuming no FIR radiation.
            Hence, information on the accurate FIR intensity is necessary
            in order to determine the gas kinetic temperature precisely,
            if $I_{160}$ is higher than 10$^2$ MJy sr$^{-1}$.
            Nevertheless, OH can still be used to explore a relative temperature structure of a cloud.
            The 1612 MHz line absorption disappears for $I_{160}$ larger than 4 $\times$ 10$^3$ MJy sr$^{-1}$,
            suggesting that the OH 18 cm transition \replaced{can not}{{cannot}} be used
            as a thermometer for clouds with such strong FIR radiation fields.
        \subsubsection{Dependence on H$_2$ density}
            Figure $\ref{fig:SEC_N15_nH2}$ shows the calculated intensities of the OH 18 cm transition
            as a function of the H$_2$ density,
            where contributions of the FIR radiation in panels (a)--(f)
            are the same as those in Figure $\ref{fig:SEC_N15_Tk}$.
            Here, the OH column density, the gas kinetic temperature, and the H$_2$ ortho-to-para ratio
            are assumed to be 3 $\times$ 10$^{15}$ cm$^{-2}$, 20 K, and 0, respectively.
            According to Figures $\ref{fig:SEC_N15_nH2}$ (e) and (f),
            the absorption of the 1720 MHz line is reproduced for an H$_2$ density lower than
            10 and 10$^3$ cm$^{-3}$, respectively.
            The H$_2$ density is required to be sufficiently low for FIR pumping to be effective,
            because collisional excitations with H$_2$ become dominant over the FIR pumping
            as the H$_2$ density increases.
            When the effect of the line overlaps is not considered (Figure $\ref{fig:SEC_N15_nH2}$ (c)),
            the 1720 MHz line is actually much weaker than the 1612 MHz line,
            but it shows a faint emission.
            When the effect of line overlaps and the FIR radiation
            assuming the gray-body are considered (Figure $\ref{fig:SEC_N15_nH2}$ (e)),
            the 1612, 1665, and 1667 MHz lines also show absorption for the H$_2$ density lower than 10 cm$^{-3}$.
            On the other hand, the absorption only for the 1720 MHz line
            is reproduced for H$_2$ densities between 10$^2$ and 10$^3$ cm$^{-3}$,
            by the DustEM FIR model, assuming the line overlaps (Figure $\ref{fig:SEC_N15_nH2}$ (f)).

\begin{figure}[htbp] 
  \centering
  \epsscale{0.95}
  \plotone{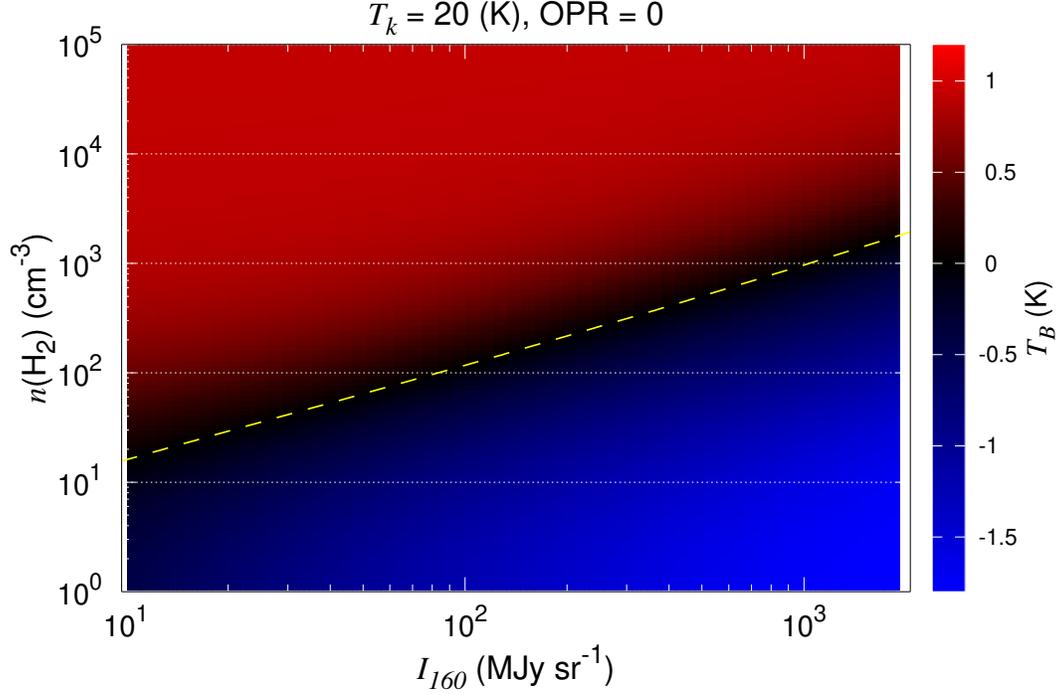}
  \caption{%
    Intensity of the 1720 MHz line on the $I_{160}$--$n$(H$_2$) plane
    derived from our statistical equilibrium calculation considering the effect of the FIR radiation
    and line overlaps.
    $I_{160}$ is used as a parameter to scale the FIR intensity profile of the DustEM model (Figure $\ref{fig:SED_fit}$).
    \replaced{Blue}{{A blue}} colored area corresponds to negative brightness temperatures of the 1720 MHz line.
    The yellow dashed line represents a 1720 MHz line intensity of 0 K.
    The FIR intensity at 160 $\mu$m ($I_{160}$) stronger than $\sim$ 10$^2$ MJy sr$^{-1}$
    is necessary to reproduce the 1720 MHz line absorption with the H$_2$ density of 10$^2$ cm$^{-3}$.
    $I_{160}$ stronger than 10$^{3}$ MJy sr$^{-1}$ is needed with the H$_2$ density of 10$^3$ cm$^{-3}$.
    \label{fig:Ndust_nH2}}
\end{figure}

            Figure $\ref{fig:Ndust_nH2}$ shows the intensity of the 1720 MHz line
            in the $n$(H$_2$)--$I_{160}$ plane.
            Here, the OH column density, the gas kinetic temperature, and the H$_2$ ortho-to-para ratio
            are assumed to be 3 $\times$ 10$^{15}$ cm$^{-2}$, 20 K, and 0, respectively (fiducial model).
            The DustEM model is employed for including the FIR radiation.
            The yellow dotted line represents the boundary between
            emission (red color) and absorption (blue color) of the 1720 MHz line,
            revealing that the threshold value of the H$_2$ density increases
            with stronger FIR radiation (Figure $\ref{fig:Ndust_nH2}$).
            This trend is reasonable, as the FIR pumping effect must be dominant over
            the collisional excitations with H$_2$ in order to produce the 1720 MHz line absorption.
            According to Figure $\ref{fig:Ndust_nH2}$,
            the absorption feature of the 1720 MHz line requires
            an H$_2$ density of $\sim$10$^3$ cm$^{-3}$ or lower for $I_{160}$
            of 10$^3$ MJy sr$^{-1}$,
            whereas it requires
            an H$_2$ density of $\sim$10$^2$ cm$^{-3}$ or lower for $I_{160}$
            of 10$^2$ MJy sr$^{-1}$.

        \subsubsection{Dependence on H$_2$ ortho-to-para ratio}
\begin{figure}[htbp] 
  \plotone{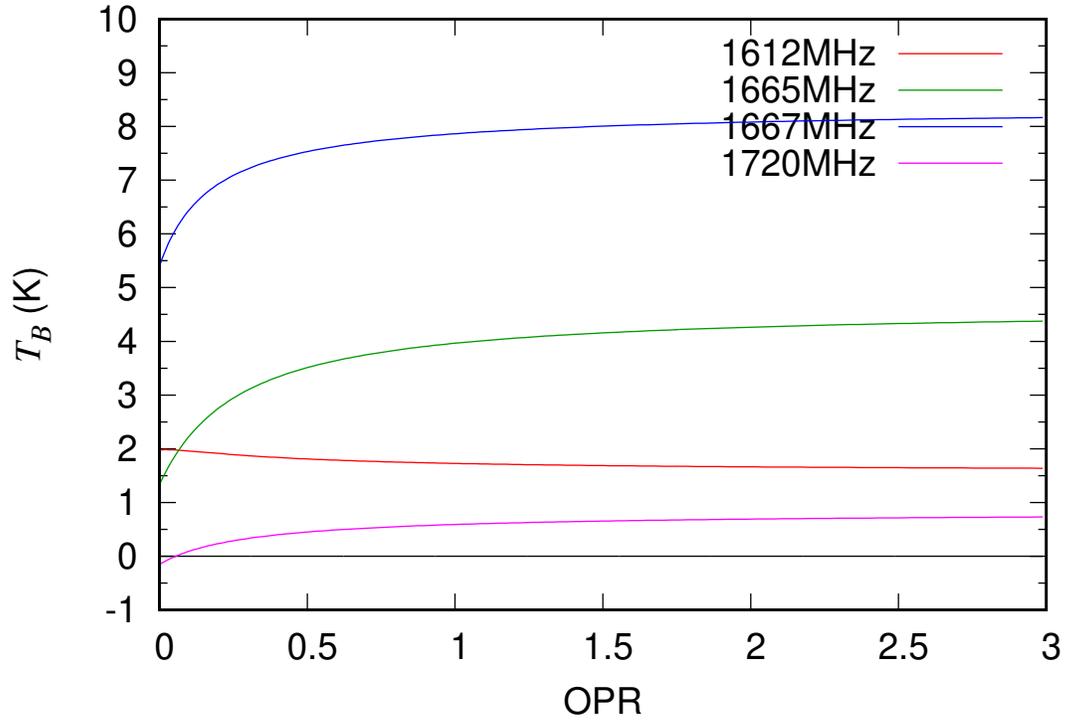}
  \caption{%
      Intensities of the hfs lines of the OH 18 cm transition
      as a function of the H$_2$ ortho-to-para ratio,
    where the H$_2$ density, the gas kinetic temperature and the column density of OH are assumed to be 10$^3$ cm$^{-3}$, 20 K, and 3 $\times$ 10$^{15}$ cm$^{-2}$, respectively.
    \label{fig:SEC_OPR}
  }
\end{figure}

            Figure $\ref{fig:SEC_OPR}$ shows the calculated intensities of the OH 18 cm hfs lines
            as a function of the H$_2$ ortho-to-para ratio for a gas kinetic temperature of 20 K.
            An H$_2$ density and the column density of OH are assumed to be 10$^3$ cm$^{-3}$ and 3 $\times$ 10$^{15}$ cm$^{-2}$, respectively.
            The ISRF strength and the H column density for the DustEM model are assumed to be
            0.4 $G_0$ and 3 $\times$ 10$^{23}$ cm$^{-2}$, respectively.
            The effect of line overlaps is considered.
            According to Figure $\ref{fig:SEC_OPR}$, the 1720 MHz line gradually becomes weaker as the H$_2$ ortho-to-para ratio becomes lower.
            The H$_2$ ortho-to-para ratio of $\sim$ 0 is necessary to reproduce the 1720 MHz line absorption, if the H$_2$ density is 10$^3$ cm$^{-3}$.
        \subsubsection{Summary of this section}\label{sec:sec_summary}
            According to our new statistical equilibrium calculations that include effects of the FIR pumping
            and the line overlaps, the absorption feature of the 1720 MHz line can be reproduced
            when the OH column density is high ($>$ 10$^{15}$ cm$^{-2}$)
            , the gas kinetic temperature is low ($<$ 30 K),
            and the H$_2$ ortho-to-para ratio is low ($\sim$0) (Figure $\ref{fig:LVGcolormap}$, left).
            The H$_2$ density is also required to be lower than 10$^3$ cm$^{-3}$ when
            the ISRF strength and the H column density are assumed to be
            0.4 $G_0$ and 3 $\times$ 10$^{23}$ cm$^{-2}$,
            respectively, in the DustEM model \citep{Jones2013}.
            The upper limit of the H$_2$ density becomes higher for stronger FIR radiation.
            On the other hand, effects of the FIR pumping and the line overlaps are negligible
            for an OH column density lower than $\sim$ 10$^{15}$ cm$^{-2}$
            and FIR intensity at 160 $\mu$m lower than $\sim$ 10$^2$ MJy sr$^{-1}$.
            The absorption feature of the 1612 MHz line disappears
            with the FIR intensity at 160 $\mu$m higher than 4 $\times$ 10$^3$ MJy sr$^{-1}$.
            Nevertheless, the OH 18 cm transition can still be used as a thermometer of a diffuse warm gas,
            as suggested by \citet{Ebisawa2015}, unless such a strong FIR field
            ($I_{160}$ $>$ 5 $\times$ 10$^3$ MJy sr$^{-1}$) is present.

            The absorption feature of the 1720 MHz line is also reproduced with
            a high column density ($>$ 10$^{15}$ cm$^{-2}$),
            a high gas kinetic temperature ($>$ 40 K),
            and the H$_2$ ortho-to-para ratio of 3 (Figure $\ref{fig:LVGcolormap}$, right).
            However, the 1612 MHz line emission is too strong compared to the observed intensities in this case.

            We have employed the collisional rate coefficients of OH reported by \citet{Offer1994} in our
            statistical equilibrium calculations.
            Recently, \citet{Klos2017} reported new collisional rate coefficients,
            although the hyperfine structure of OH was not considered.
            We obtained hfs-resolved rate coefficients by private communication (from Dr. Fran\c{c}ois Lique),
            which are approximately calculated from these fine structure rate coefficients by using the M-J random method \citep{Klos2017}.
            In order to examine the robustness of our statistical equilibrium calculations described
            in the previous subsections,
            we also conducted the calculations by using these collisional rate coefficients.
            The results are presented in Appendix B.
            Even with these collisional rate coefficients,
            the essential features of our results on the hfs intensity anomaly do not change significantly.
            Thus, uncertainties of the collisional rate coefficients are considered
            to have a limited effect on the results of the hfs intensity anomalies of the OH 18 cm transition.

    \subsection{Fitting the observed spectra}
        We compare our new model \replaced{to}{{with}} the spectra of the OH 18 cm transition
        observed toward TMC-1FN, TMC-1(CP), L183, and L169.
        Since the 8.2 km s$^{-1}$ and 6.5 km s$^{-1}$ components of TMC-1FN
        and the $\sim$ 6.4 km s$^{-1}$ component of TMC-1 (CP)
        show absorption or faint emission in the 1612 MHz line,
        they are expected to trace warm molecular gas \citep{Ebisawa2015}.
        Hence, we determined the gas kinetic temperature and the OH column density
        for these components (Table $\ref{table:SECresult}$)
        assuming an H$_2$ density of 10$^3$ cm$^{-3}$.
        On the other hand, the 5.5 km s$^{-1}$ component of TMC-1FN,
        the $\sim$ 5.4 km s$^{-1}$ component of TMC-1 (CP),
        L183, and L169
        show absorption or faint emission of the 1720 MHz line,
        and hence, they likely trace cold dense regions.

        \subsubsection{TMC-1FN}\label{sec:TMC-1FN}
\begin{figure}[htbp] 
  \centering
  \epsscale{0.5}
  \plotone{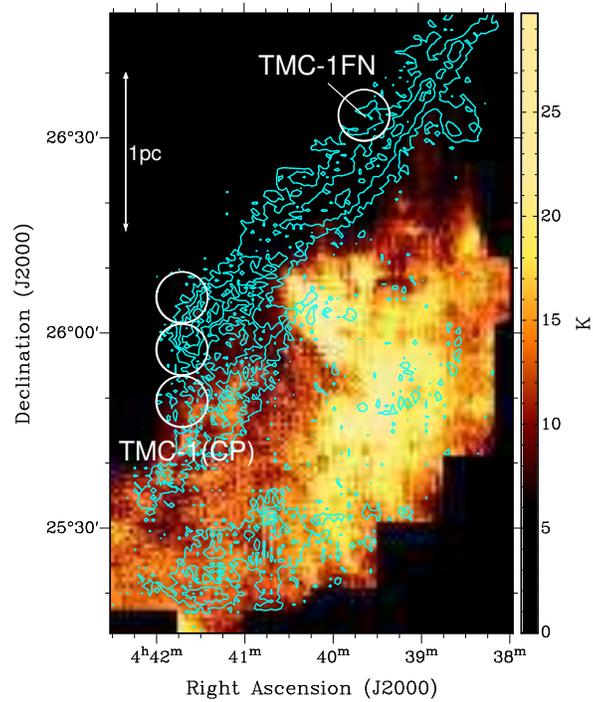}
  \caption{%
    Integrated intensity map of C$^{18}$O ($J$=1--0)
    (5.15 km s$^{-1}$ $< V_{LSR} <$ 5.45 km s$^{-1}$; contours)
    and {}$^{13}$CO ($J$=1--0)
    (6.45 km s$^{-1}$ $< V_{LSR} <$ 6.95 km s$^{-1}$; color)
    observed toward HCL2 \citep{Sunada1999}.
    Circles represent the positions observed in OH,
    and their diameter corresponds to the FWHM of the Effelsberg 100 m telescope.
    \label{fig:HCL2}}
\end{figure}

            The 1720 MHz line absorption is detected with $\sim$3 $\sigma$ confidence level toward TMC-1FN.
            To explain this feature,
            the FIR intensity of $\sim$ 10$^3$ MJy sr$^{-1}$ at 160 $\mu$m is necessary according to our new statistical equilibrium calculations.
            This is stronger by one order of magnitude than the value observed with {\it Spitzer}
            (80--120 MJy sr$^{-1}$) according to \citet{Flagey2009}.
            In order to explain this, we assess the geometrical effect of a filamentary structure,
            for which the FIR intensity inside might be enhanced.
            Assuming that the filament is a cylinder with a Plummer-like density profile \citep{Malinen2012},
            the FIR intensity inside the filament can be three to four times stronger than the observed value.
            Details of the calculation are described in Appendix C.
            A further factor of two to three necessary to explain the 1720 MHz absorption might be explained by contributions from
            a filamentary structure in the northwest of TMC-1FN \citep{Xu2016a},
            another filament near TMC-1 (CP) (Figure $\ref{fig:HCL2}$),
            or by diffuse gas extended in the southwestern part of HCL2 (Figure $\ref{fig:HCL2}$).
            However, contributions from other filamentary structures might be small compared to
            the contribution from the nearest filament, as they occupy a smaller solid angle
            than the nearest filament.
            In addition, contributions from the extended gas component are limited,
            considering that the observed FIR intensity of this component is comparable
            to that observed in TMC-1FN.
            Approximations employed in the calculations---in particular,
            neglect of the velocity gradient in the treatment of the line overlap \citep{Guilloteau1981}---
            may also contribute to this discrepancy.
            Contributions from non-local line overlaps \citep{Cesaroni1991, Elitzur1976a}
            caused by the velocity gradient within the cloud
            might explain more efficient FIR pumping,
            although quantitative analysis on this effect is rather difficult.

            For the 8.2 km s$^{-1}$ component, the gas kinetic temperature and the OH column density are determined to be 70 $\pm$ 3 K and (1.4 $\pm$ 0.4) $\times$ 10$^{14}$ cm$^{-2}$,
            respectively, where the H$_2$ density and the ortho-to-para ratio are assumed to be 10$^3$ cm$^{-3}$ and 3, respectively.
            For the 6.5 km s$^{-1}$ component, the gas kinetic temperature and the OH column density are determined to be 30 $\pm$ 3 K and (3.6 $\pm$ 0.3) $\times$ 10$^{14}$ cm$^{-2}$, respectively.
            An H$_2$ ortho-to-para ratio of 0.1 provides the best fit, although it cannot be constrained by the fitting procedure.
            For the 5.5 km s$^{-1}$ component,
            the gas kinetic temperature and the OH column density are estimated to be lower than 30 K,
            and higher than 10$^{15}$ cm$^{-2}$, respectively,
            in order to reproduce the 1720 MHz line absorption according to our statistical equilibrium analysis.

\begin{figure}[htbp] 
  \centering
  \epsscale{1}
  \includegraphics[width=0.8\hsize]{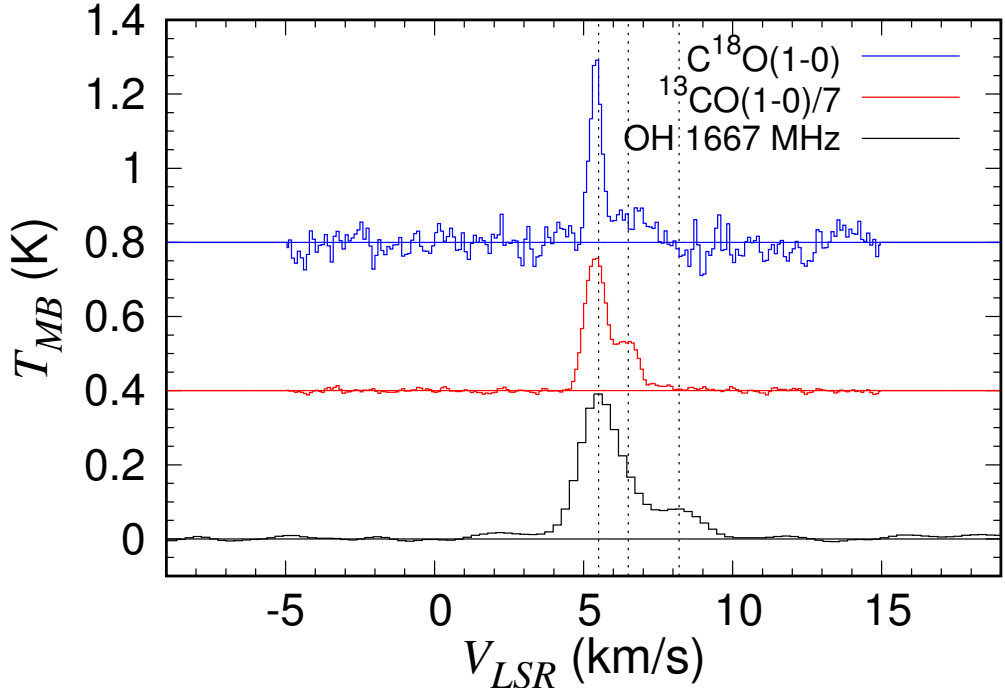}
  \caption{
    Spectra for the 1667 MHz (black) hfs line of the OH 18 cm transition,
    the $^{13}$CO ($J=1-0$) line (red)
    and
    the C$^{18}$O ($J=1-0$) line (blue) observed toward TMC-1FN (Figure $\ref{fig:HCL2}$).
    The $^{13}$CO and C$^{18}$O lines are averaged over the beam size of the OH 18 cm transition,
    as revealed by the white circle in Figure $\ref{fig:HCL2}$.
    Dashed lines show the velocities of the three components (5.5 km s$^{-1}$, 6.5 km s$^{-1}$, and 8.2 km s$^{-1}$).
    \label{fig:sTMC1FNCO}
  }
\end{figure}

            Figure $\ref{fig:sTMC1FNCO}$ shows the spectrum of the 1667 MHz line of OH observed toward
            TMC-1FN\replaced{, overlaid on}{{with}} the $^{13}$CO ($J=1-0$) and C$^{18}$O ($J=1-0$) line spectra observed toward the same position with NRO 45-m telescope \citep{Sunada1999}.
            The $^{13}$CO and C$^{18}$O line profiles are prepared by averaging these lines' spectra over the beam size of the OH 18 cm transition.
            As shown in Figure $\ref{fig:sTMC1FNCO}$, the 8.2 km s$^{-1}$ component, which is clearly seen in OH, is not detected in both $^{13}$CO and C$^{18}$O.
            On the other hand, the 6.5 km s$^{-1}$ component is clearly seen in $^{13}$CO and C$^{18}$O,
            while the C$^{18}$O emission is prominent for the 5.5 km s$^{-1}$ component.
            These results suggest that the 8.2 km s$^{-1}$ component traces diffuse warm gas that cannot readily be traced by $^{13}$CO,
            whereas the 5.5 km s$^{-1}$ component traces relatively cold and dense gas also traced by C$^{18}$O.
            The 6.5 km s$^{-1}$ component appears to trace gas at intermediate temperature and density between these two structures.
            This picture is consistent with the derived gas kinetic temperatures and the H$_2$ ortho-to-para ratio (Table $\ref{table:SECresult}$).
            In fact, the gas kinetic temperature should be higher in the outer part of the cloud,
            which produces a deeper absorption feature of the 1612 MHz line and brighter 1720 MHz emission.
            Since the higher $V_{LSR}$ components tend to trace a warmer part of the cloud in TMC-1FN (Table $\ref{table:SECresult}$),
            the observed spectra of OH seem to reflect the cloud structure with the increasing gas kinetic temperature toward cloud peripheries as does the increasing $V_{LSR}$.

\begin{figure}[htbp] 
  \centering
  \epsscale{1}
  \includegraphics[width=0.5\hsize]{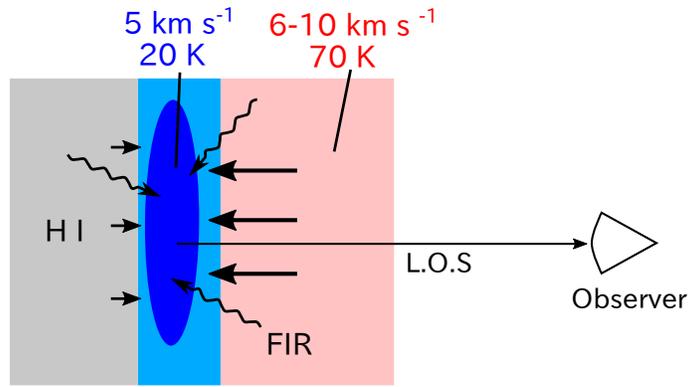}
  \caption{
    Schematic illustration of the cross section of TMC-1FN along the line-of-sight.
    The cold ($\sim$ 20 K) and a dense core (5 km s$^{-1}$) is formed by compressive motion of the extended warm gas (6--10 km s$^{-1}$).
    We observe the source from the right hand side.
    \label{fig:TMC1FN_illust}
  }
\end{figure}

            Figure $\ref{fig:HCL2}$ shows the integrated intensity maps of C$^{18}$O ($J$=1--0) (cyan contours) and $^{13}$CO ($J$=1--0) (color) observed with the NRO 45-m telescope for the entire region of HCL2 \citep{Sunada1999}.
            A prominent straight structure is seen in the northern part of HCL2 in the C$^{18}$O map,
            which also has been pointed out by \citet{Narayanan2008} and \citet{Goldsmith2010}.
            Interestingly, the velocity of the blue-shifted component detected in the OH 18 cm transition ($V_{LSR} \sim 5.5$ km s$^{-1}$) corresponds to that of C$^{18}$O exhibiting the straight structure,
            whereas the velocity of the red-shifted component seen in OH ($V_{LSR} \sim$ 6--10 km s$^{-1}$) corresponds to that of the southwestern extended gas in the $^{13}$CO map.
            These two structures are interfacing with each other, suggesting a compressive motion as the origin of the straight structure (Figure $\ref{fig:TMC1FN_illust}$).
            Such a compressive motion is consistent with the decreasing velocity from southwest to northeast across the straight structure observed in $^{13}$CO and C$^{18}$O.
            In addition, the existence of an [H I] cloud in the northeast of TMC-1FN is suggested by the archival data of The Galactic Arecibo L-band Feed Array HI (GALFA-HI) Survey maps \citep{Peek2011}.
            Furthermore, the 6--10 km s$^{-1}$ component traces a diffuse warm gas,
            whereas the 5.5 km s$^{-1}$ component traces a cold and dense core illuminated by the FIR radiation from dust grains, according to our statistical equilibrium calculation.
            It is likely that the 5.5 km s$^{-1}$ component traces cold cores surrounded by the warm envelope gas in HCL2.

            The 6.5 and 8.2 km s$^{-1}$ components show the 1612 MHz line absorption,
            in spite of the FIR pumping effect seen in the 5.5 km s$^{-1}$ component.
            This can be explained by the above picture of TMC-1FN (Figure $\ref{fig:TMC1FN_illust}$).
            A warm envelope gas traced by the 6.5 and 8.2 km s$^{-1}$ components
            is likely to be illuminated by fainter FIR radiation
            than the gas traced by the 5.5 km s$^{-1}$ component.
            The OH column density for the 6.5 and 8.2 km s$^{-1}$ components is also lower.
            According to the results in Appendix C,
            the FIR intensity at 160 $\mu$m is estimated to be $\sim$400 and 100 MJy sr$^{-1}$
            at positions 0.5 and 1.0 pc distant from the center of the TMC-1FN filament, respectively.
            The OH column densities are determined to be
            3.6 $\times$ 10$^{14}$ cm$^{-2}$ and 1.4 $\times$ 10$^{14}$ cm$^{-2}$
            for the 6.5 and 8.2 km s$^{-1}$ components, respectively (Table $\ref{table:SECresult}$).
            As described in Section $\ref{sec:sec_summary}$,
            our calculation shows that the FIR pumping effect is negligible
            for an OH column density lower than 10$^{15}$ cm$^{-2}$ and $I_{160}$ lower than 100 MJy sr$^{-1}$.
            On the other hand, the 1612 MHz line indeed appears in absorption for $I_{160}$ lower than 4 $\times$ 10$^3$ MJy sr$^{-1}$,
            if the OH column density is 3 $\times$ 10$^{14}$ cm$^{-2}$ (Figure $\ref{fig:LVG_FIR}$ (d)).
            Hence, even though there might be a \replaced{strong}{{moderate}} FIR field
            \added{{($\sim$ 10$^3$ MJy sr$^{-1}$)}},
            the 1612 MHz line absorption in the 6.5 and 8.2 km s$^{-1}$ components can be reproduced
            as long as the OH column density is lower than $\sim$ 10$^{15}$ cm$^{-2}$\replaced{, whereas}{{. In this case,}}
            the intensities of the 1612 and 1720 MHz lines
            \replaced{might be increased and decreased by a factor of 2--3, respectively}
            {{are slightly affected by the FIR pumping effect}} (Figure $\ref{fig:LVG_FIR}$).

        \subsubsection{TMC-1 (CP)}
            The OH spectra toward TMC-1 (CP) consists of the two components (5.36--5.59 km s$^{-1}$ and 6.34--6.62 km s$^{-1}$), as mentioned in Section $\ref{sec:Result}$.
            The gas kinetic temperature is determined to be about 40 K for the red-shifted component
            (6.34--6.62 km s$^{-1}$ component),
            which shows weak emission or absorption of the 1612 MHz line.
            On the other hand, the gas kinetic temperature is considered to be lower
            than 30 K for the blue-shifted component (5.36--5.59 km s$^{-1}$),
            according to the DustEM model considering the effect of the line overlaps.
            The FIR intensity is assumed to be 1.4 $\times$ 10$^3$ MJy sr$^{-1}$ at 160 $\mu$m.
            However, only an upper limit to the gas kinetic temperature can be determined,
            because of the limited signal-to-noise ratio.
            The derived parameters for the red-shifted components are summarized in Table $\ref{table:SECresult}$.

        \subsubsection{L183 and L169}
            As shown in Figure $\ref{fig:sL183}$, the 1720 MHz line shows weak emission in L183, and it is totally absent in L169.
            \replaced{The}{{From these results, the}} gas kinetic temperature is
            \replaced{required}{{estimated}} to be lower than 30 K according to our statistical equilibrium calculations that include FIR pumping (Figure $\ref{fig:LVGcolormap}$).
            The gas kinetic temperature of 30 K is lower than values determined for positions 8$'$ and 16$'$ south from L183 in our previous paper \citep{Ebisawa2015} (33 K and 57 K, respectively).
            Previous studies have shown that there is a dense core in the northern part of L183, which is traced in the NH$_3$ line (cross position in Figure $\ref{fig:L183}$) \citep{Swade1989, Dickens2000}.
            A rather diffuse molecular cloud seen in {}$^{13}$CO emission is extended toward its southern part \citep{Laureijs1995, Lehtinen2003}.
            Thus the increase of the gas kinetic temperature toward the southern peripheries in L183 indicates that photoelectric heating by interstellar UV radiation is less effective in the core position.
\section{SUMMARY}\label{sec:summary}
    We have shown that the absorption feature of the 1720 MHz line of the OH 18 cm transition can be reproduced
    by taking account of the effect of pumping of FIR radiation from dust grains.
    We have developed a statistical equilibrium calculation code considering the effects of the FIR radiation and line overlaps, and have investigated the origin of the 1720 MHz line absorption.
    As a result, the 1720 MHz absorption is found to trace a dense ($N$(OH) $>$ 10$^{15}$ cm$^{-2}$) and cold ($T_k$ $<$ 30 K) core illuminated by relatively strong FIR radiation from dust grains in surrounding clouds.
    On the other hand, we have confirmed that the absorption feature of the 1612 MHz line can be used to determine the gas kinetic temperature of the diffuse warm gas \replaced{over}{{for}} a wide range of \added{{an}} H$_2$ density,
    even in the presence of the FIR radiation.
    Toward TMC-1FN and TMC-1(CP), the gas kinetic temperature is determined to be 30--70 K for the red-shifted component,
    whereas it is lower than $\sim$ 30 K for the blue-shifted component,
    in which the 1720 MHz line is seen in absorption.
    Combined analyses of OH, $^{13}$CO, and C$^{18}$O suggest that the peculiar straight structure in HCL2 is likely formed by the compression of the warm envelope gas in southwestern direction.
    Hence, its unique characteristics, combined with the excitation calculations, make the quartet of the hfs lines of the OH 18 cm transition
    a new probe to study molecular cloud formation and a tracer of CO-dark molecular gas.

    Nevertheless, we still have a shortage of the FIR intensity by a factor of two to three
    in order to fully reproduce the 1720 MHz absorption,
    even if we consider the filamentary structure of the cloud.
    This contradiction might be explained by contributions from extended gas and other filaments,
    and effect of non-local line overlaps.
    Since quantitative analyses of these effects are rather difficult,
    they will be explored in future works.

    However, this study has clarified the mechanism of the observed 1720 MHz absorption
    in molecular clouds and its physical meaning,
    and we believe that it will be a stiff base for such future studies.
    A full understanding of the hyperfine anomalies of the OH 18 cm transition is
    essentially important for future observational studies of OH
    using the Five-hundred-meter Aperture Spherical Radio Telescope (FAST)
    and Square Kilometre Array (SKA).

\acknowledgements
We are grateful to Dr. Fran\c{c}ois Lique for providing us new hfs-resolved collisional rates
based on the M-J random method prior to publication.
We are also grateful to the anonymous reviewer for invaluable comments.
We thank Mr. Hiroshi Inokuma for his assistance in the early stage of this work.
We thank the staff of the Effelsberg 100 m telescope of MPIfR for their excellent support.
We also thank Prof. Takashi Onaka for helpful discussion on the interstellar FIR field.
This study is supported by Grants-in-Aid from Ministry of Education, Sports, Science, and Technologies of Japan (25400223, 25108005, 18H05222, and 17J02168).


\bibliography{cites}

\begin{thebibliography}{}
\expandafter\ifx\csname natexlab\endcsname\relax\def\natexlab#1{#1}\fi

\bibitem[{{Boulanger} {et~al.}(1996){Boulanger}, {Abergel}, {Bernard},
  {Burton}, {Desert}, {Hartmann}, {Lagache}, \& {Puget}}]{Boulanger1996}
{Boulanger}, F., {Abergel}, A., {Bernard}, J.-P., {et~al.} 1996, \aap, 312, 256

\bibitem[{{Caswell}(2004)}]{Caswell2004}
{Caswell}, J.~L. 2004, \mnras, 349, 99

\bibitem[{{Cesaroni} \& {Walmsley}(1991)}]{Cesaroni1991}
{Cesaroni}, R., \& {Walmsley}, C.~M. 1991, \aap, 241, 537

\bibitem[{{Cohen}(1995)}]{Cohen1995}
{Cohen}, R.~J. 1995, \apss, 224, 55

\bibitem[{{Compi{\`e}gne} {et~al.}(2011){Compi{\`e}gne}, {Verstraete}, {Jones},
  {Bernard}, {Boulanger}, {Flagey}, {Le Bourlot}, {Paradis}, \&
  {Ysard}}]{Compiegene2011}
{Compi{\`e}gne}, M., {Verstraete}, L., {Jones}, A., {et~al.} 2011, \aap, 525,
  A103

\bibitem[{{Darling} \& {Giovanelli}(2002)}]{Darling2002}
{Darling}, J., \& {Giovanelli}, R. 2002, \aj, 124, 100

\bibitem[{{Dickens} {et~al.}(2000){Dickens}, {Irvine}, {Snell}, {Bergin},
  {Schloerb}, {Pratap}, \& {Miralles}}]{Dickens2000}
{Dickens}, J.~E., {Irvine}, W.~M., {Snell}, R.~L., {et~al.} 2000, \apj, 542,
  870

\bibitem[{{Draine} \& {Katz}(1986)}]{Draine1986}
{Draine}, B.~T., \& {Katz}, N. 1986, \apj, 306, 655

\bibitem[{{Ebisawa} {et~al.}(2015){Ebisawa}, {Inokuma}, {Sakai}, {Menten},
  {Maezawa}, \& {Yamamoto}}]{Ebisawa2015}
{Ebisawa}, Y., {Inokuma}, H., {Sakai}, N., {et~al.} 2015, \apj, 815, 13

\bibitem[{{Elitzur}(1976)}]{Elitzur1976b}
{Elitzur}, M. 1976, \apj, 203, 124

\bibitem[{{Elitzur} {et~al.}(1976){Elitzur}, {Goldreich}, \&
  {Scoville}}]{Elitzur1976a}
{Elitzur}, M., {Goldreich}, P., \& {Scoville}, N. 1976, \apj, 205, 384

\bibitem[{{Flagey} {et~al.}(2009){Flagey}, {Noriega-Crespo}, {Boulanger},
  {Carey}, {Brooke}, {Falgarone}, {Huard}, {McCabe}, {Miville-Desch{\^e}nes},
  {Padgett}, {Paladini}, \& {Rebull}}]{Flagey2009}
{Flagey}, N., {Noriega-Crespo}, A., {Boulanger}, F., {et~al.} 2009, \apj, 701,
  1450

\bibitem[{{Goldsmith} {et~al.}(2010){Goldsmith}, {Velusamy}, {Li}, \&
  {Langer}}]{Goldsmith2010}
{Goldsmith}, P.~F., {Velusamy}, T., {Li}, D., \& {Langer}, W.~D. 2010, \apj,
  715, 1370

\bibitem[{{Guilloteau} {et~al.}(1981){Guilloteau}, {Lucas}, \&
  {Omont}}]{Guilloteau1981}
{Guilloteau}, S., {Lucas}, R., \& {Omont}, A. 1981, \aap, 97, 347

\bibitem[{{Gundermann} {et~al.}(1965){Gundermann}, {Goldstein}, \&
  {Lilley}}]{Gundermann1965}
{Gundermann}, E.~J., {Goldstein}, Jr., S.~J., \& {Lilley}, A.~E. 1965, \aj, 70,
  321

\bibitem[{{Habing}(1968)}]{Habing1968}
{Habing}, H.~J. 1968, \bain, 19, 421

\bibitem[{{Harju} {et~al.}(2000){Harju}, {Winnberg}, \&
  {Wouterloot}}]{Harju2000}
{Harju}, J., {Winnberg}, A., \& {Wouterloot}, J.~G.~A. 2000, \aap, 353, 1065

\bibitem[{{Hoffman} {et~al.}(2005){Hoffman}, {Goss}, {Brogan}, \&
  {Claussen}}]{Hoffman2005}
{Hoffman}, I.~M., {Goss}, W.~M., {Brogan}, C.~L., \& {Claussen}, M.~J. 2005,
  \apj, 620, 257

\bibitem[{{Jones} {et~al.}(2013){Jones}, {Fanciullo}, {K{\"o}hler},
  {Verstraete}, {Guillet}, {Bocchio}, \& {Ysard}}]{Jones2013}
{Jones}, A.~P., {Fanciullo}, L., {K{\"o}hler}, M., {et~al.} 2013, \aap, 558,
  A62

\bibitem[{{Kanekar} {et~al.}(2004){Kanekar}, {Chengalur}, \&
  {Ghosh}}]{Kanekar2004}
{Kanekar}, N., {Chengalur}, J.~N., \& {Ghosh}, T. 2004, Physical Review
  Letters, 93, 051302

\bibitem[{{K\l os} {et~al.}(2017){K\l os}, Ma, {Dagdigian}, {Alexander},
  {Faure}, \& {Lique}}]{Klos2017}
{K\l os}, J., Ma, Q., {Dagdigian}, P.~J., {et~al.} 2017, Monthly Notices of the
  Royal Astronomical Society, 471, 4249

\bibitem[{{Laureijs} {et~al.}(1995){Laureijs}, {Fukui}, {Helou}, {Mizuno},
  {Imaoka}, \& {Clark}}]{Laureijs1995}
{Laureijs}, R.~J., {Fukui}, Y., {Helou}, G., {et~al.} 1995, \apjs, 101, 87

\bibitem[{{Lehtinen} {et~al.}(2003){Lehtinen}, {Mattila}, {Lemke}, {Juvela},
  {Prusti}, \& {Laureijs}}]{Lehtinen2003}
{Lehtinen}, K., {Mattila}, K., {Lemke}, D., {et~al.} 2003, \aap, 398, 571

\bibitem[{{Malinen} {et~al.}(2012){Malinen}, {Juvela}, {Rawlings},
  {Ward-Thompson}, {Palmeirim}, \& {Andr{\'e}}}]{Malinen2012}
{Malinen}, J., {Juvela}, M., {Rawlings}, M.~G., {et~al.} 2012, \aap, 544, A50

\bibitem[{{Narayanan} {et~al.}(2008){Narayanan}, {Heyer}, {Brunt}, {Goldsmith},
  {Snell}, \& {Li}}]{Narayanan2008}
{Narayanan}, G., {Heyer}, M.~H., {Brunt}, C., {et~al.} 2008, \apjs, 177, 341

\bibitem[{{Offer} {et~al.}(1994){Offer}, {van Hemert}, \& {van
  Dishoeck}}]{Offer1994}
{Offer}, A.~R., {van Hemert}, M.~C., \& {van Dishoeck}, E.~F. 1994, \jcp, 100,
  362

\bibitem[{{Peek} {et~al.}(2011){Peek}, {Heiles}, {Douglas}, {Lee}, {Grcevich},
  {Stanimirovi{\'c}}, {Putman}, {Korpela}, {Gibson}, {Begum}, {Saul},
  {Robishaw}, \& {Kr{\v c}o}}]{Peek2011}
{Peek}, J.~E.~G., {Heiles}, C., {Douglas}, K.~A., {et~al.} 2011, \apjs, 194, 20

\bibitem[{{Plummer}(1911)}]{Plummer1911}
{Plummer}, H.~C. 1911, \mnras, 71, 460

\bibitem[{{Sunada} \& {Kitamura}(1999)}]{Sunada1999}
{Sunada}, K., \& {Kitamura}, Y. 1999, in Interstellar Turbulence, ed.
  J.~{Franco} \& A.~{Carraminana}, 208

\bibitem[{{Swade}(1989)}]{Swade1989}
{Swade}, D.~A. 1989, \apj, 345, 828

\bibitem[{{van Langevelde} {et~al.}(1995){van Langevelde}, {van Dishoeck},
  {Sevenster}, \& {Israel}}]{Langevelde1995}
{van Langevelde}, H.~J., {van Dishoeck}, E.~F., {Sevenster}, M.~N., \&
  {Israel}, F.~P. 1995, \apjl, 448, L123

\bibitem[{{Weaver} {et~al.}(1965){Weaver}, {Williams}, {Dieter}, \&
  {Lum}}]{Weaver1965}
{Weaver}, H., {Williams}, D.~R.~W., {Dieter}, N.~H., \& {Lum}, W.~T. 1965,
  \nat, 208, 29

\bibitem[{{Weisberg} {et~al.}(2005){Weisberg}, {Johnston}, {Koribalski}, \&
  {Stanimirovi{\'c}}}]{Weisberg2005}
{Weisberg}, J.~M., {Johnston}, S., {Koribalski}, B., \& {Stanimirovi{\'c}}, S.
  2005, Science, 309, 106

\bibitem[{{Wiesemeyer} {et~al.}(2012){Wiesemeyer}, {G{\"u}sten}, {Heyminck},
  {Jacobs}, {Menten}, {Neufeld}, {Requena-Torres}, \&
  {Stutzki}}]{Wiesemeyer2012}
{Wiesemeyer}, H., {G{\"u}sten}, R., {Heyminck}, S., {et~al.} 2012, \aap, 542,
  L7

\bibitem[{{Wolak} {et~al.}(2012){Wolak}, {Szymczak}, \&
  {G{\'e}rard}}]{Wolak2012}
{Wolak}, P., {Szymczak}, M., \& {G{\'e}rard}, E. 2012, \aap, 537, A5

\bibitem[{{Xu} {et~al.}(2016){Xu}, {Li}, {Yue}, \& {Goldsmith}}]{Xu2016a}
{Xu}, D., {Li}, D., {Yue}, N., \& {Goldsmith}, P.~F. 2016, \apj, 819, 22

\end{thebibliography}
%
%
\clearpage
\begin{rotatetable} 
\begin{deluxetable}{lllllllll}
  \tabletypesize{\scriptsize}
  \tablewidth{0pt}
  \tablecaption{
    Observed Line Parameters
    \label{table:linepara}}
  \tablehead{%
    Source                   & R.A.                  & Decl.                      & $V_{LSR}$     & $\Delta V$    & $T_{1612}$ & $T_{1665}$ & $T_{1667}$ & $T_{1720}$ \\
                             & (2000)                & (2000)                     & (km s$^{-1}$) & (km s$^{-1}$) & (K)        & (K)        & (K)        & (K)          }
    \startdata
    TMC-1CP(0$'$, 24$'$)     & 04$^h$41$^m$42$^s$.88 & +26$^{\circ}$05$'$27$''$.0 & 5.44(2)       & 1.55(5)       & 0.14(1)    & 0.29(1)    & 0.50(1)    & -0.01(1)   \\
                             &                       &                            & 6.62(30)      & 3.67(39)      & 0.009(8)   & 0.067(12)  & 0.054(13)  & 0.045(9)   \\ \hline
    TMC-1CP(0$'$, 16$'$)     & 04$^h$41$^m$42$^s$.88 & +25$^{\circ}$57$'$27$''$.0 & 5.36(2)       & 1.47(8)       & 0.12(1)    & 0.24(2)    & 0.44(2)    & -0.025(16) \\
                             &                       &                            & 6.35(21)      & 3.37(26)      & -0.011(10) & 0.11(1)    & 0.11(2)    & 0.042(10)  \\ \hline
    TMC-1CP(0$'$, 8$'$)      & 04$^h$41$^m$42$^s$.88 & +25$^{\circ}$49$'$27$''$.0 & 5.59(2)       & 1.65(8)       & 0.16(1)    & 0.27(3)    & 0.49(3)    & -0.049(22) \\
                             &                       &                            & 6.34(15)      & 3.07(18)      & -0.029(13) & 0.15(2)    & 0.16(3)    & 0.088 (15) \\ \hline
    TMC-1FN                  & 04$^h$39$^m$37$^s$.66 & +26$^{\circ}$33$'$27$''$.0 & 5.46(2)       & 1.41(3)       & 0.12(1)    & 0.20(1)    & 0.36(1)    & -0.032(7)  \\
    (TMC-1CP(-28$'$, 52$'$)) &                       &                            & 6.51(5)       & 1.54(11)      & 0.000(7)   & 0.15(1)    & 0.087(13)  & 0.068(5)   \\
                             &                       &                            & 8.16(7)       & 1.85(15)      & -0.047(4)  & 0.026(4)   & 0.076(4)   & 0.026(4)   \\ \hline
    L183                     & 15$^h$54$^m$00$^s$.54 & -02$^{\circ}$51$'$48$''$.5 & 2.47(1)       & 1.37(2)       & 0.12(1)    & 0.41(1)    & 0.68(1)    & 0.073(8)   \\ \hline
    L169                     & 15$^h$51$^m$00$^s$.21 & -03$^{\circ}$26$'$48$''$.5 & 3.53(2)       & 1.70(5)       & 0.079(6)   & 0.15(1)    & 0.27(1)    & 0.001(6)   \\ \hline
    \enddata
    \tablewidth{500pt}
  \tablecomments{%
    Intensities of the four hfs components of OH ($T_{1612}, T_{1665}, T_{1667}, T_{1720}$), linewidth ($\Delta V$) and $V_{LSR}$ are obtained by a Gaussian fit,
    assuming that the $\Delta V$ and $V_{LSR}$ values of the four hfs components are the same for \replaced{the same}{{each}} observed position.
    The error in the parentheses represents one standard deviation.
   }
\end{deluxetable}
\end{rotatetable} 
\clearpage
\begin{deluxetable}{ccc} 
  \tabletypesize{\small}
  \tablewidth{0pt}
  \tablecaption{
    Velocity separation between the hyperfine structure components of the
    $J=5/2$, $^2\Pi_{3/2}$ -- $J=3/2$, $^2\Pi_{3/2}$
    transitions of OH (top),
    and that between hfs components of the
    $J=1/2$, $^2\Pi_{1/2}$ -- $J=3/2$, $^2\Pi_{3/2}$ transitions (bottom).
    Levels in the first and second columns are represented in Figure
    $\ref{fig:OH_energy_FIR}$.
    \label{table:overlap}}
  \tablehead{%
    transition 1     & transition 2     & d$V$\\
    (level1, level2) & (level1, level2) & (km/s)
    }
  \startdata
     (6, 2)  & (5, 2)  & 1.6 \\
     (6, 2)  & (5, 1)  & 4.7 \\
     (5, 2)  & (5, 1)  & 6.3 \\
     (8, 4)  & (7, 4)  & 2.1 \\
     (8, 4)  & (7, 3)  & 4.4 \\
     (7, 4)  & (7, 3)  & 6.5 \\ \hline
     (10, 2) & (10, 1) & 4.2 \\
     (10, 2) & (9, 1)  & 3.0 \\
     (10, 1) & (9, 1)  & 1.2 \\
     (12, 4) & (12, 3) & 4.4 \\
     (12, 4) & (11, 3) & 2.8 \\
     (12, 3) & (11, 3) & 7.1 \\ \hline
  \enddata
\end{deluxetable} 
\clearpage
\begin{deluxetable}{lccccc} 
  \tabletypesize{\small}
  \tablewidth{0pt}
  \tablecaption{
    Best fit parameters determined by our statistical equilibrium calculation.
    \label{table:SECresult}}
  \tablehead{%
    Source               & $V_{LSR}$ & $n$(H$_2$)    & $T_k$ & $N$(OH)                    & o/p    \\
                         & (km/s)    & (cm$^{-3}$)   & (K)   & (cm$^{-2}$)                &     }
    \startdata
    TMC-1CP(0$'$, 24$'$) & 6.62      & 10$^3$(fixed) & 31(1) & 5.1(14) $\times$ 10$^{14}$ & 0.1(fixed) \\
    TMC-1CP(0$'$, 16$'$) & 6.35      & 10$^3$(fixed) & 35(1) & 5.0(4)  $\times$ 10$^{14}$ & 0.3(fixed) \\
    TMC-1CP(0$'$, 8$'$)  & 6.34      & 10$^3$(fixed) & 43(1) & 5.6(3)  $\times$ 10$^{14}$ & 0.7(fixed) \\ \hline
    TMC-1FN              & 6.51      & 10$^3$(fixed) & 30(3) & 3.6(3)  $\times$ 10$^{14}$ & 0.1(fixed) \\
                         & 8.16      & 10$^3$(fixed) & 70(3) & 1.4(4)  $\times$ 10$^{14}$ & 3.0(fixed) \\ \hline
    \enddata
  \tablecomments{%
   We derived the OH column density and the gas kinetic temperature for the 6.5 km s$^{-1}$ and the 8.2 km s$^{-1}$ components of TMC-1FN, and the $\sim$ 6.4 km s$^{-1}$ component of TMC-1 (CP),
   where the H$_2$ ortho-to-para ratio has been changed from 0 to 3 with a 0.05 interval, and is fixed to the value that gives the best-fit result.
   The error in the parentheses represents one standard deviation.
   The $V_{LSR}$ corresponds to those in Table $\ref{table:linepara}$.
   }
\end{deluxetable} 
\clearpage
\clearpage
\clearpage
\clearpage
\clearpage
\clearpage
\clearpage
\clearpage
\clearpage
\clearpage
\clearpage
\clearpage
\clearpage
\begin{figure}[htbp] 
  \centering
  \epsscale{1.05}
  \plotone{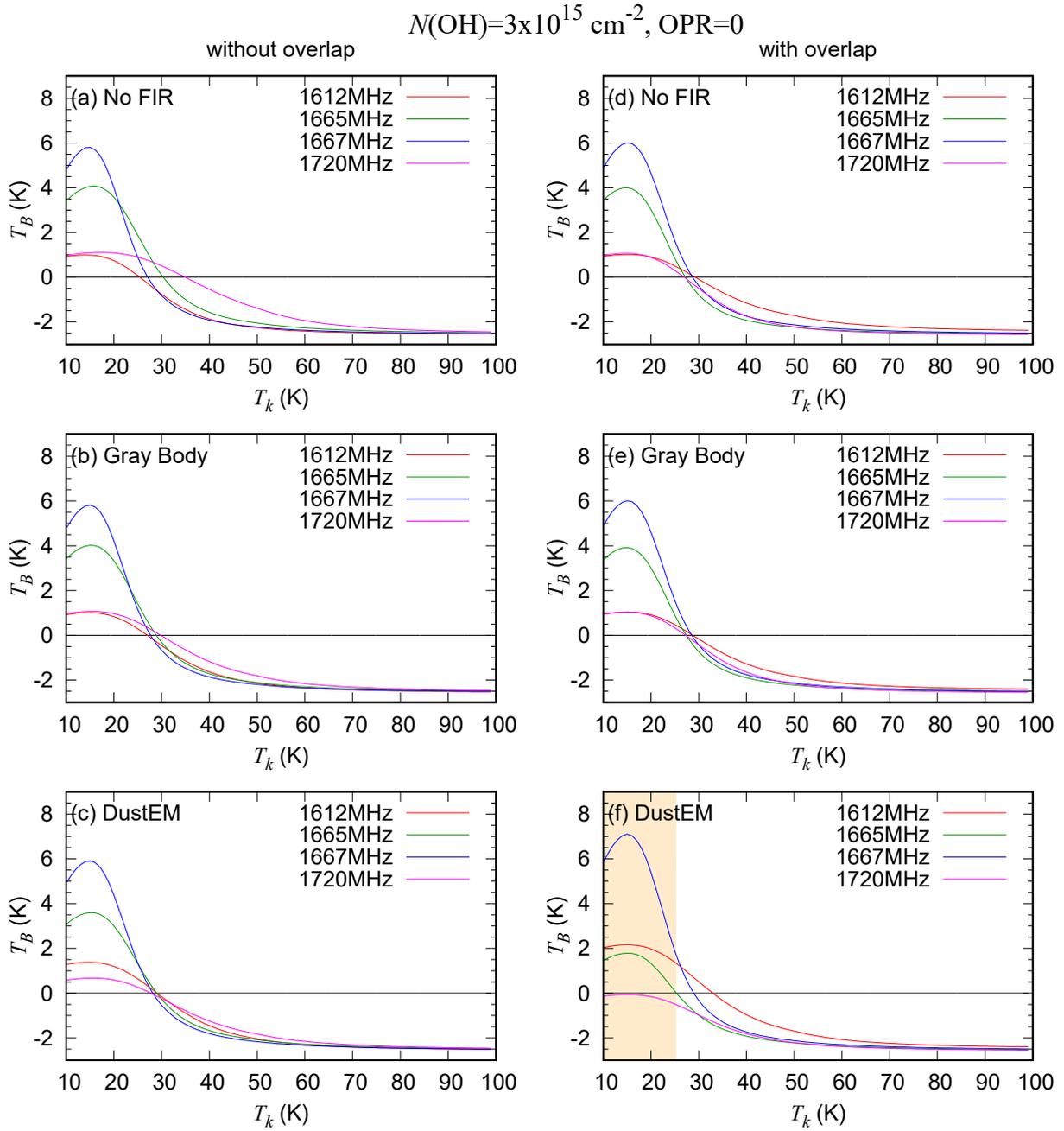}
  \caption{%
    Intensities of the hfs lines of the OH 18 cm transition
    as a function of the gas kinetic temperature,
    where the H$_2$ density, the OH column density,
    and the ortho-to-para ratio of H$_2$ are assumed to be
    10$^3$ cm$^{-3}$, 3 $\times$ 10$^{15}$ cm$^{-2}$, and 0, respectively.
    The effect of line overlaps is not considered in the left panels (a)--(c),
    whereas it is considered in the right panels (d)--(f).
    The FIR radiation is not considered in the top panels (a) and (d),
    whereas the gray-body approximation with the dust temperature of 15 K
    is applied in the middle panels (b) and (e).
    The FIR radiation calculated by the fiducial DustEM model \citep{Jones2013}
    is employed in the bottom panels (c) and (f).
    The orange area in the panel (f) represents the region where the 1720 MHz line appears in absorption and the other three lines show emission.
    \label{fig:SEC_N15_Tk}
  }
\end{figure}
\clearpage
\begin{figure}[htbp] 
  \centering
  \epsscale{1.0}
  \plotone{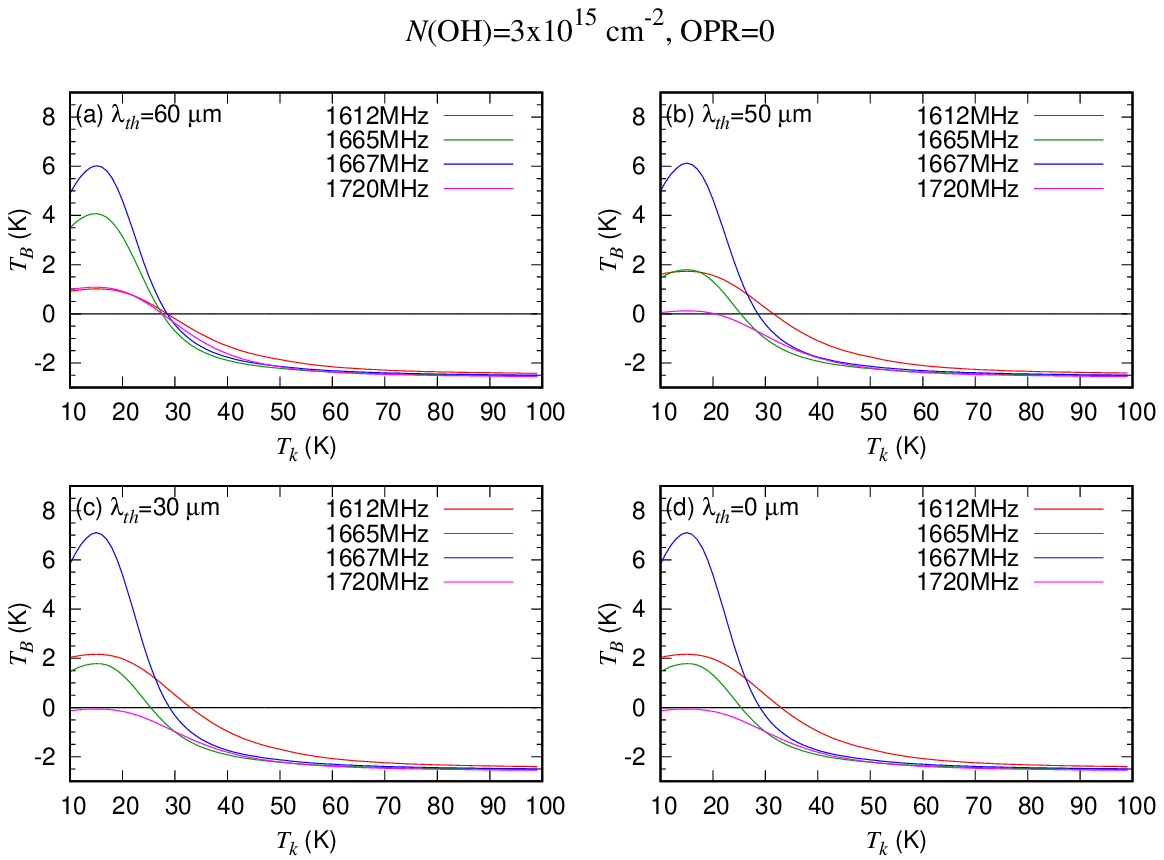}
  \caption{%
    Intensities of the hfs lines of the OH 18 cm transition
    as a function of the gas kinetic temperature,
    derived by introducing the artificial \replaced{threshold}{{thresholds}} for the
    shortest wavelength in the fiducial DustEM model \citep{Jones2013},
    where the H$_2$ density, the OH column density, and the ortho-to-para ratio
    of H$_2$ are assumed to be 10$^3$,
    3 $\times$ 10$^{15}$ cm$^{-2}$, and 0, respectively.
    Models assumed in each panel (a)--(d) are the same as those in
    Figure $\ref{fig:SEC_N15_Tk}$ (f) except for the FIR field.
    The threshold ($\lambda_{th}$) is (a) 60 $\mu$m, (b) 50 $\mu$m, (c) 30 $\mu$m.
    No FIR radiation is assumed at the wavelength shorter than the threshold.
    Panel (a) is the same as the panel (f) of Figure $\ref{fig:SEC_N15_Tk}$.
    \label{fig:SEC_N15_Tk_fcut}
  }
\end{figure}
\clearpage
\clearpage
\begin{figure}[htbp] 
  \centering
  \epsscale{1.2}
  \plotone{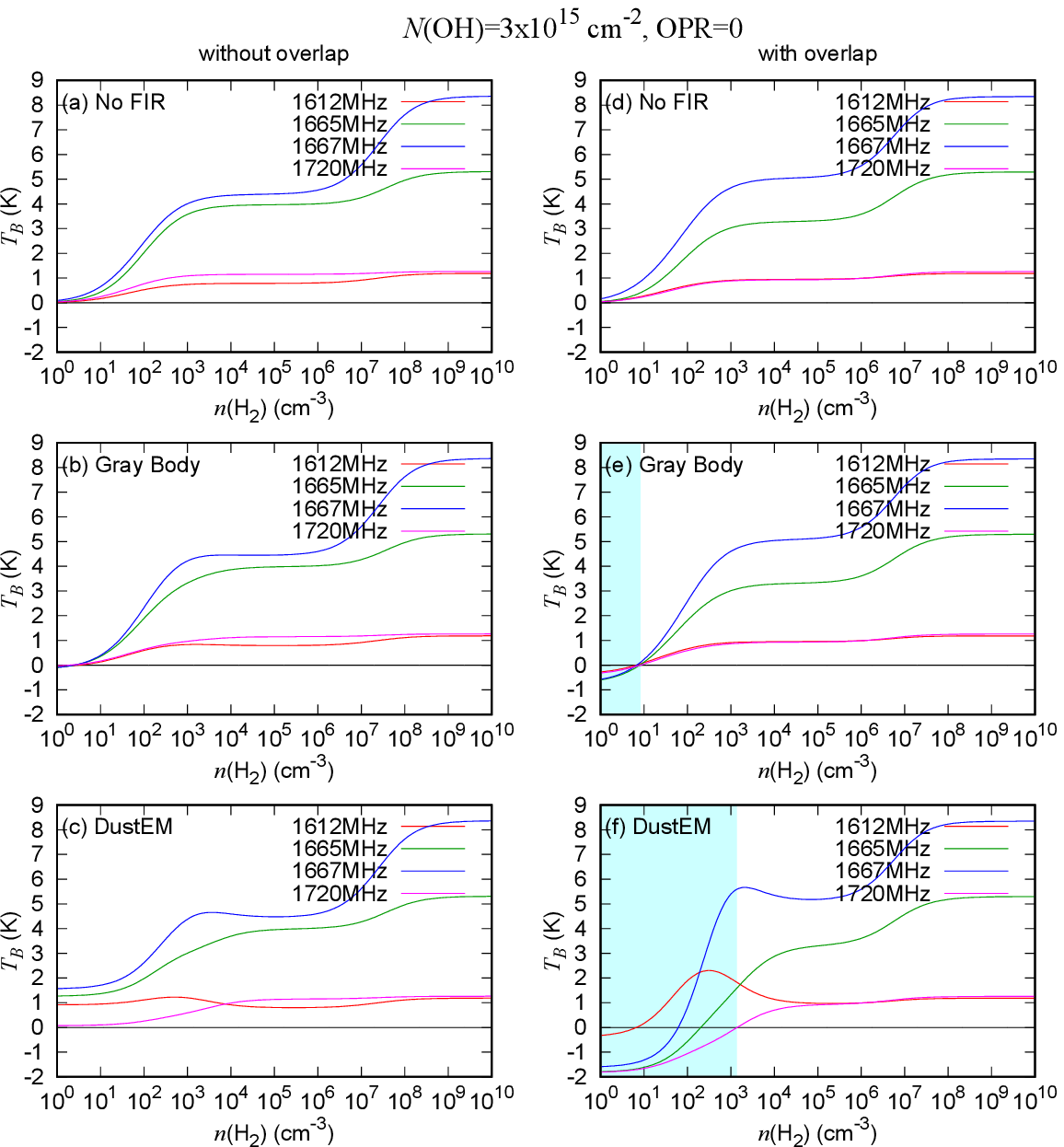}
  \caption{%
    Intensities of the hfs lines of the OH 18 cm transition
    as a function of the H$_2$ density,
    where the OH column density, the gas kinetic temperature, and the ortho-to-para ratio of H$_2$ are assumed to be
    3 $\times$ 10$^{15}$ cm$^{-2}$, 20 K, and 0, respectively.
    Models assumed in each panel (a)--(f) are the same as those in Figure $\ref{fig:SEC_N15_Tk}$.
    The cyan areas in the panels (e) and (f) represent the region where the 1720 MHz line appears in absorption.
    \label{fig:SEC_N15_nH2}
  }
\end{figure}
\clearpage
\clearpage
\clearpage
\clearpage
\clearpage
\begin{figure}[htbp] 
  \centering
  \epsscale{1.2}
  \plotone{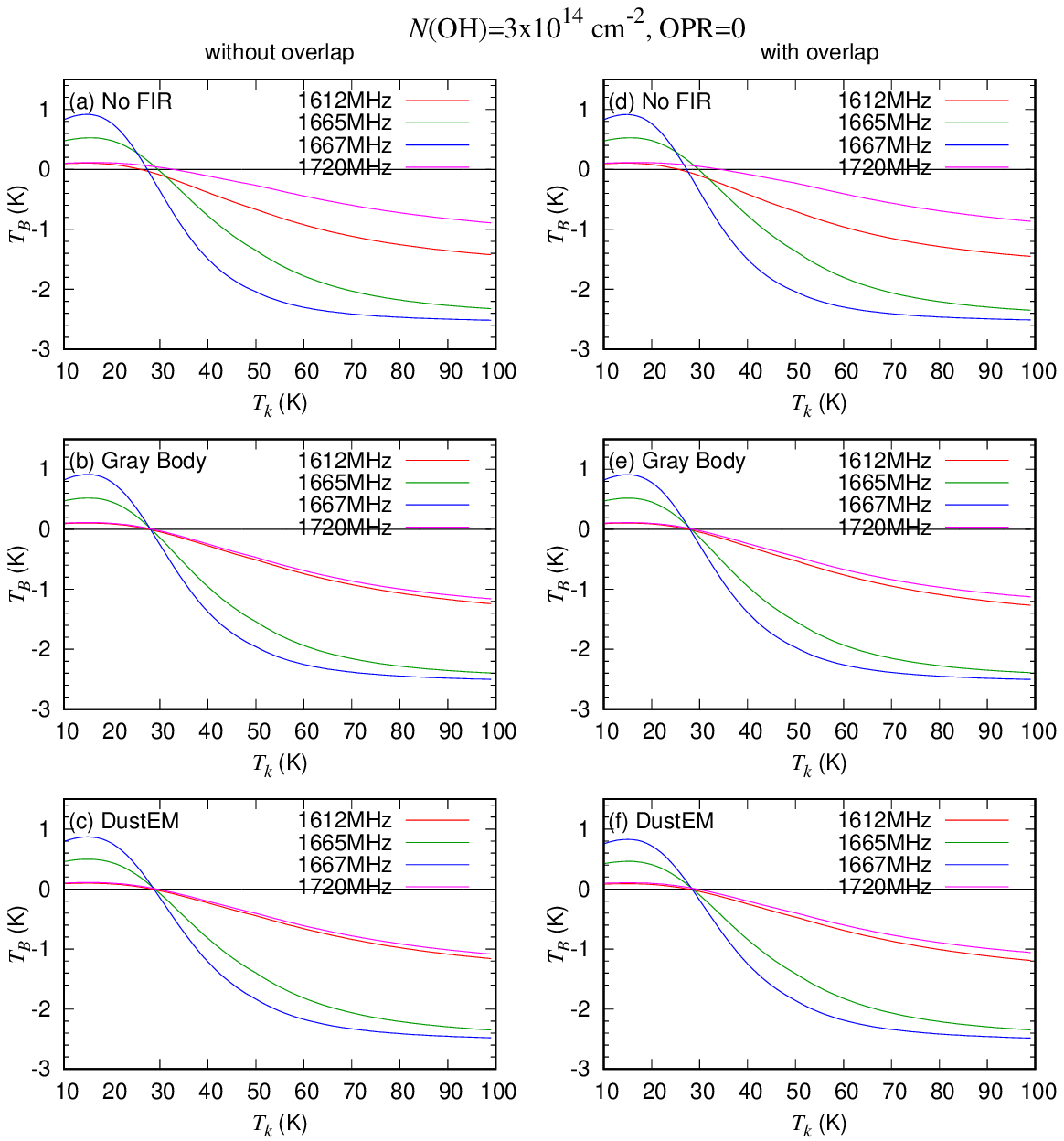}
  \caption{%
    Intensities of the hfs lines of the OH 18 cm transition
    as a function of the gas kinetic temperature,
    where the H$_2$ density, the OH column density, and the ortho-to-para ratio of H$_2$ are assumed to be
    10$^3$ cm$^{-3}$, 3 $\times$ 10$^{14}$ cm$^{-2}$, and 0, respectively.
    Models assumed in each panel (a)--(f) are the same as those in Figure $\ref{fig:SEC_N15_Tk}$.
    \label{fig:SEC_N14_Tk}
  }
\end{figure}
\clearpage
\begin{figure}[htbp] 
  \centering
  \epsscale{1.2}
  \plotone{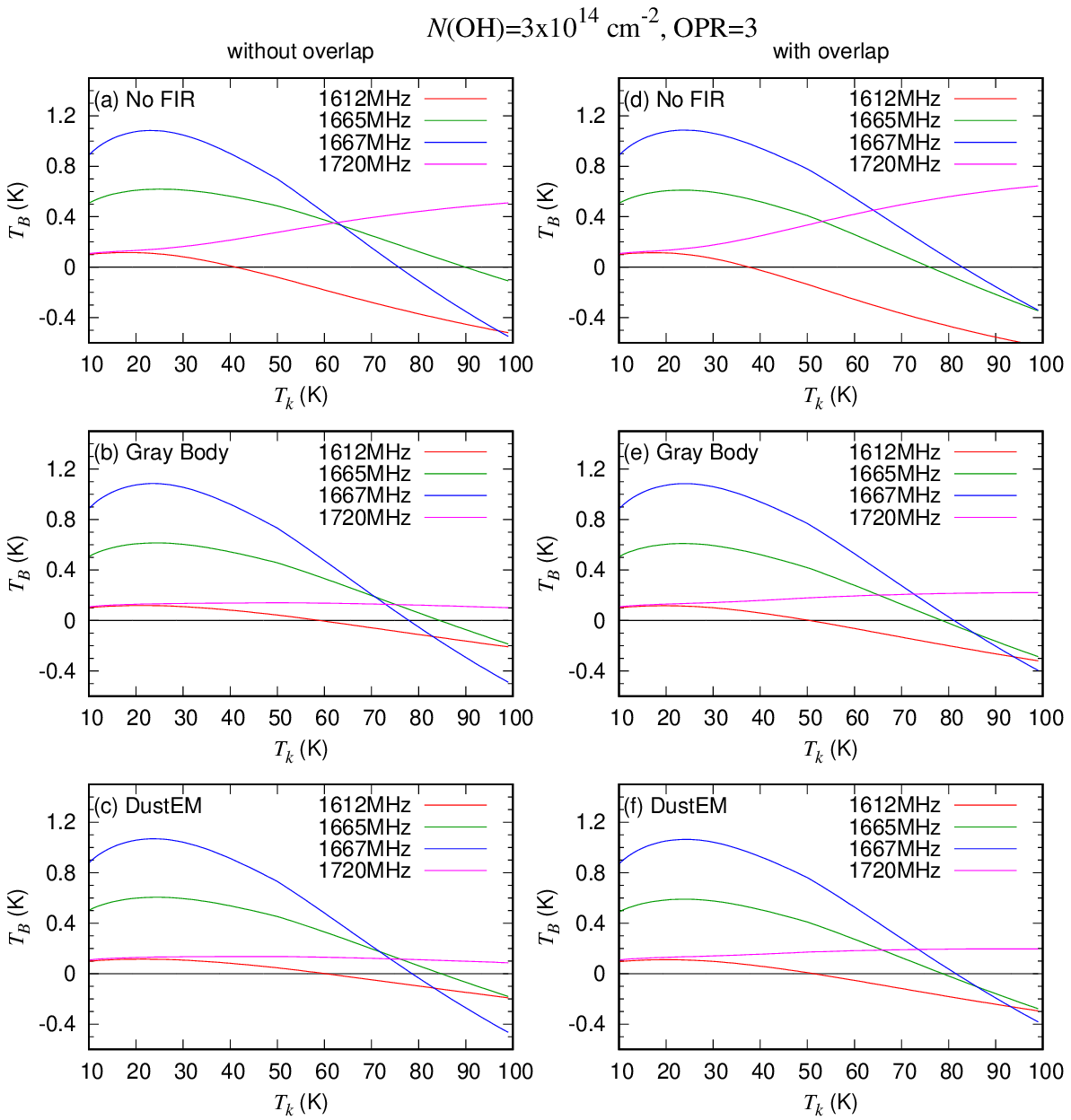}
  \caption{%
    Intensities of the hfs lines of the OH 18 cm transition
    as a function of the gas kinetic temperature,
    where the H$_2$ density, the OH column density, and the ortho-to-para ratio of H$_2$ are assumed to be
    10$^3$ cm$^{-3}$, 3 $\times$ 10$^{14}$ cm$^{-2}$, and 3, respectively.
    Models assumed in each panel (a)--(f) are the same as those in Figure $\ref{fig:SEC_N15_Tk}$.
    \label{fig:SEC_N14_OPR3_Tk}
  }
\end{figure}
\clearpage
\begin{figure}[htbp] 
  \centering
  \epsscale{1.2}
  \plotone{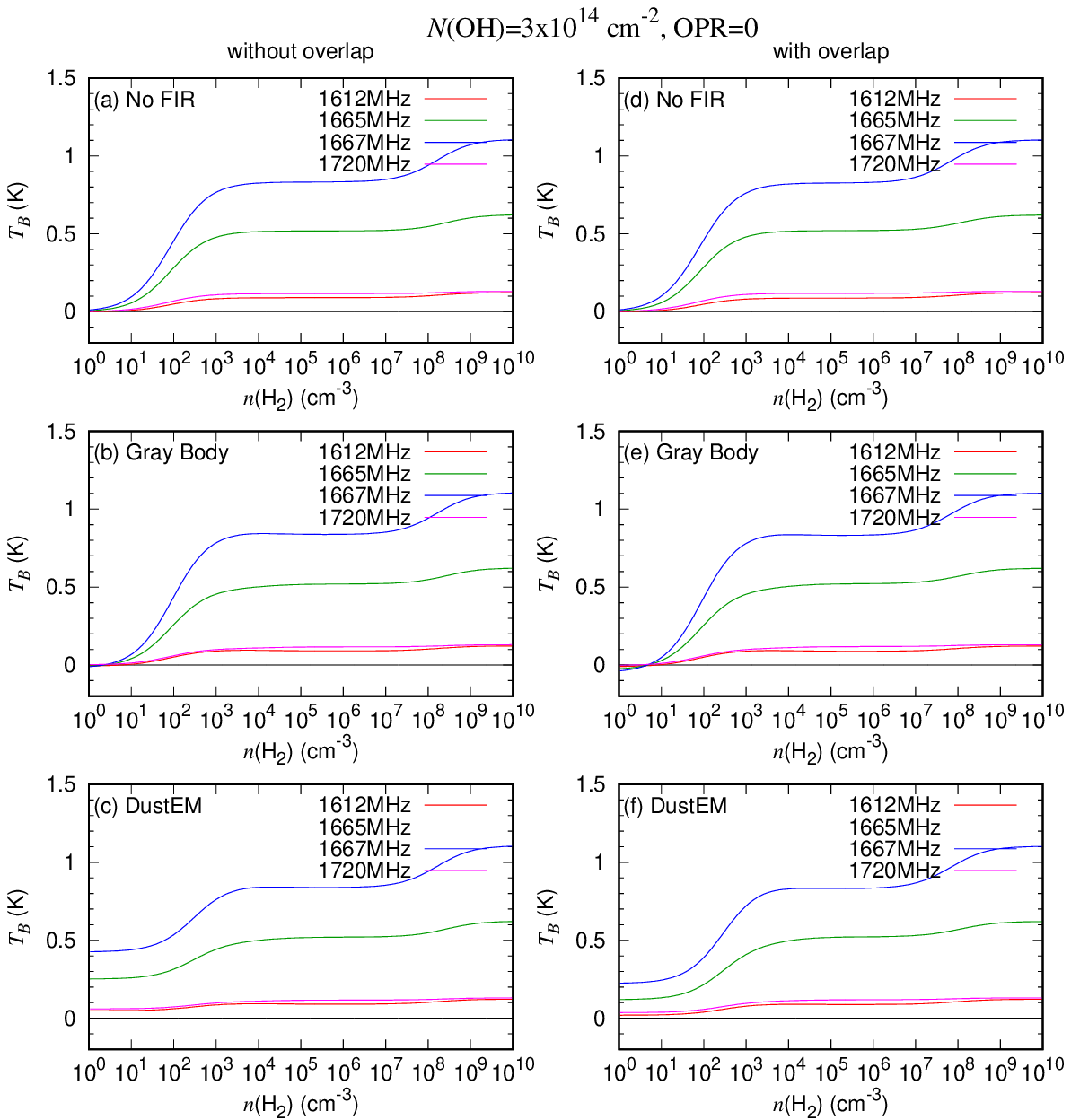}
  \caption{%
    Intensities of the hfs lines of the OH 18 cm transition
    as a function of the H$_2$ density,
    where the OH column density, the gas kinetic temperature, and the ortho-to-para ratio of H$_2$ are assumed to be
    3 $\times$ 10$^{14}$ cm$^{-2}$, 20 K, and 0, respectively.
    Models assumed in each panel (a)--(f) are the same as those in Figure $\ref{fig:SEC_N15_Tk}$.
    \label{fig:SEC_N14_nH2}
  }
\end{figure}
\clearpage
\begin{figure}[htbp] 
  \centering
  \epsscale{1.2}
  \plotone{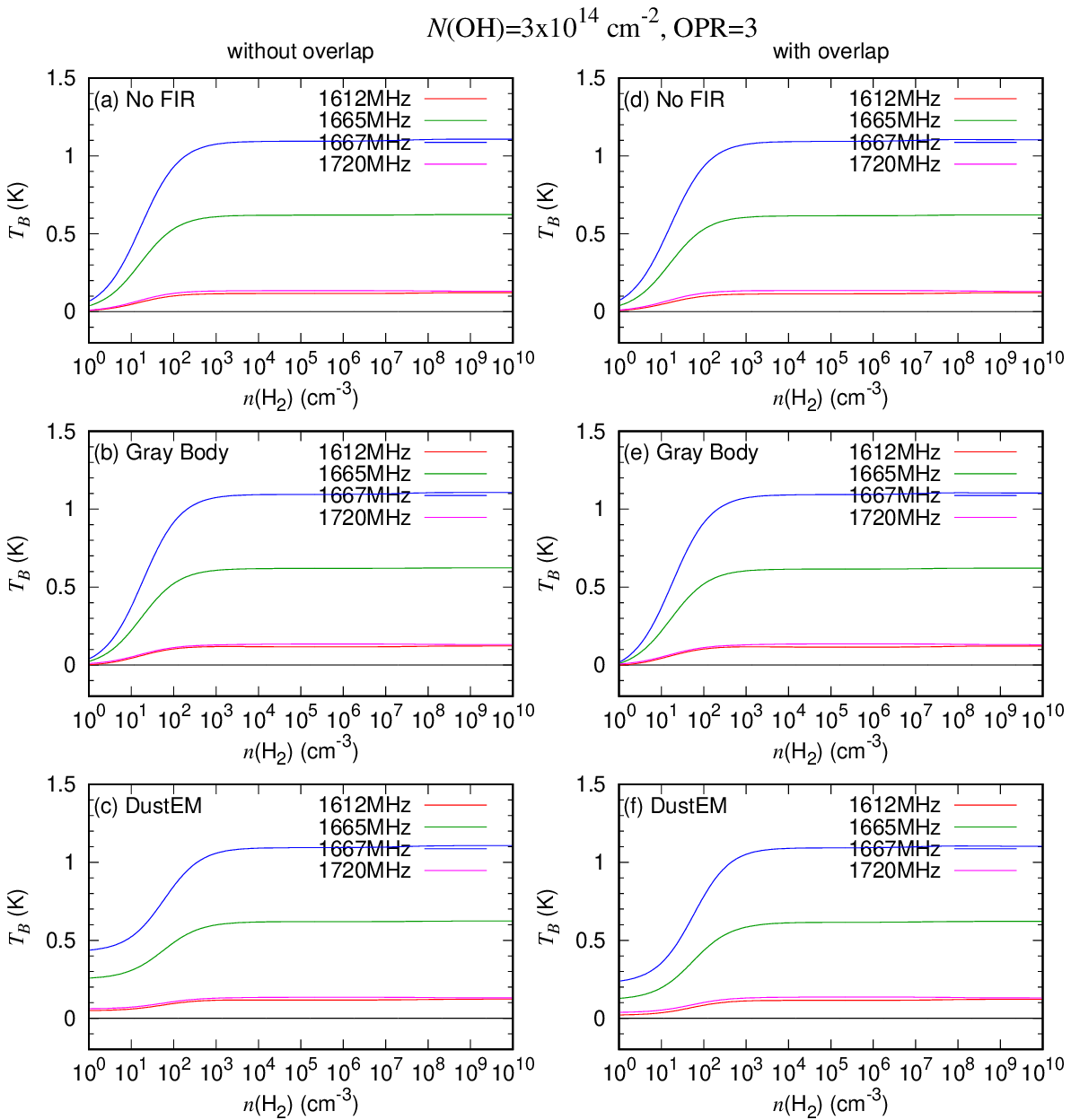}
  \caption{%
    Intensities of the hfs lines of the OH 18 cm transition
    as a function of the H$_2$ density,
    where the OH column density, the gas kinetic temperature, and the ortho-to-para ratio of H$_2$ are assumed to be
    3 $\times$ 10$^{14}$ cm$^{-2}$, 20 K, and 3, respectively.
    Models assumed in each panel (a)--(f) are the same as those in Figure $\ref{fig:SEC_N15_Tk}$.
    \label{fig:SEC_N14_OPR3_nH2}
  }
\end{figure}
\clearpage
\begin{figure}[htbp] 
  \centering
  \epsscale{1.2}
  \plotone{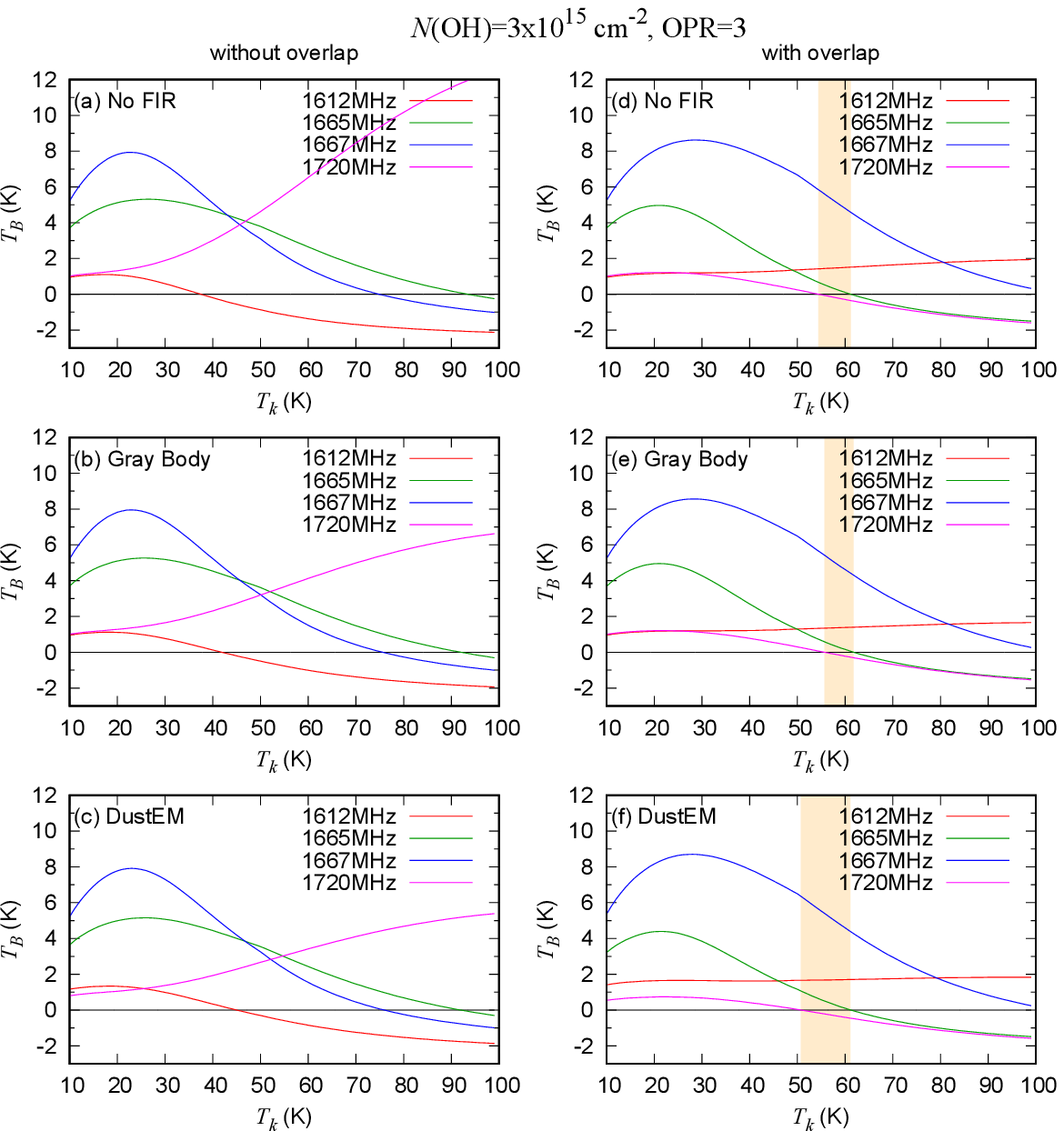}
  \caption{%
    Intensities of the hfs lines of the OH 18 cm transition
    as a function of the gas kinetic temperature,
    where the H$_2$ density, the OH column density, and the ortho-to-para ratio of H$_2$ are assumed to be
    10$^3$ cm$^{-3}$, 3 $\times$ 10$^{15}$ cm$^{-2}$, and 3, respectively.
    Models assumed in each panel (a)--(f) are the same as those in Figure $\ref{fig:SEC_N15_Tk}$.
    The orange areas in the panels (d)--(f) represent the region where the 1720 MHz line appears in absorption and the other three lines show emission.
    \label{fig:SEC_N15_OPR3_Tk}
  }
\end{figure}
\clearpage
\begin{figure}[htbp] 
  \centering
  \epsscale{1.2}
  \plotone{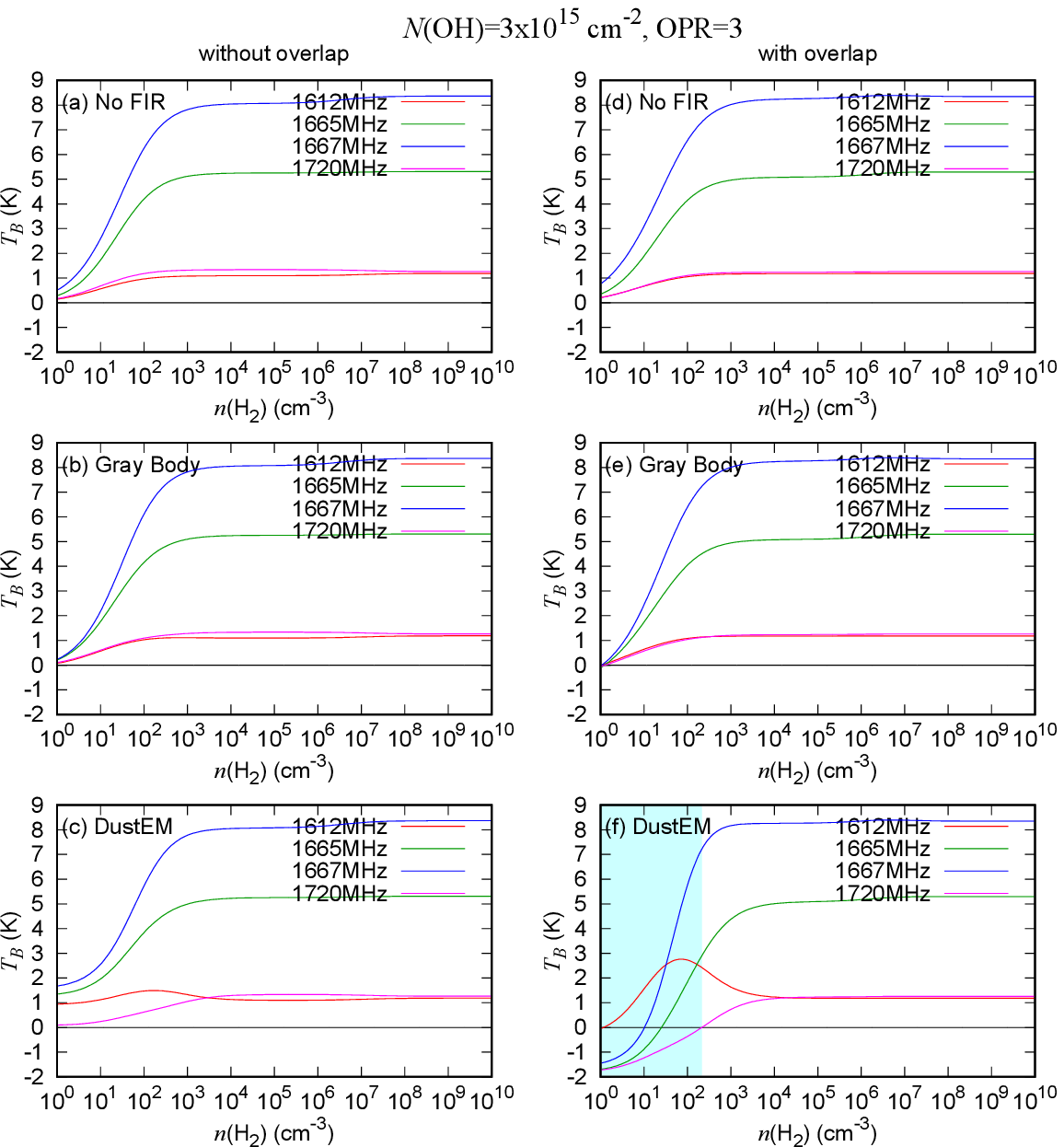}
  \caption{%
    Intensities of the hfs lines of the OH 18 cm transition
    as a function of the H$_2$ density,
    where the OH column density, the gas kinetic temperature, and the ortho-to-para ratio of H$_2$ are assumed to be
    3 $\times$ 10$^{15}$ cm$^{-2}$, 20 K, and 3, respectively.
    Models assumed in each panel (a)--(f) are the same as those in Figure $\ref{fig:SEC_N15_Tk}$.
    The cyan area in the panel (f) represents the region where the 1720 MHz line appears in absorption.
    \label{fig:SEC_N15_OPR3_nH2}
  }
\end{figure}
\clearpage
\begin{figure}[htbp] 
  \centering
  \epsscale{1.2}
  \plotone{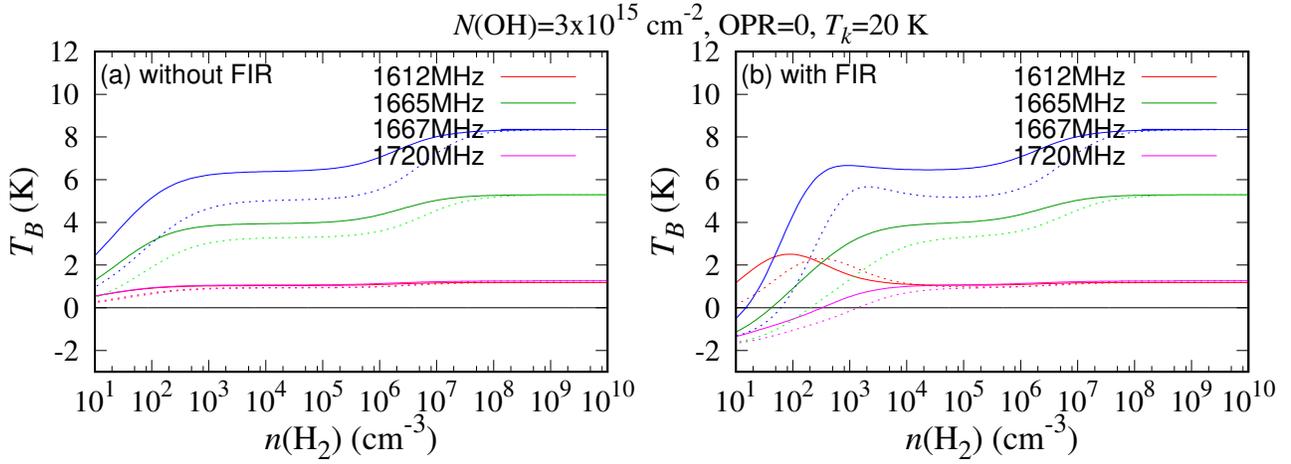}
  \caption{%
    Intensities of the hfs lines of the OH 18 cm transition
    as a function of the H$_2$ density.
    Dotted lines represent results with the collisional rate coefficients of OH calculated by \citet{Offer1994},
    whereas solid lines show those with the new rate coefficients calculated by applying M-J random method to hfs-unresolved rates by \citet{Klos2017} (solid line).
    Effect of the FIR radiation is not included in the left panel,
    whereas FIR radiation calculated with the fiducial DustEM model
    ($N$(H)=3 $\times$ 10$^{23}$ cm$^{-2}$, ISRF=0.4 $G_0$)
    is included in the right panel.
    \label{fig:newrate}}
\end{figure}
\clearpage
\begin{figure}[!htbp] 
  \centering
  \epsscale{0.8}
  \plotone{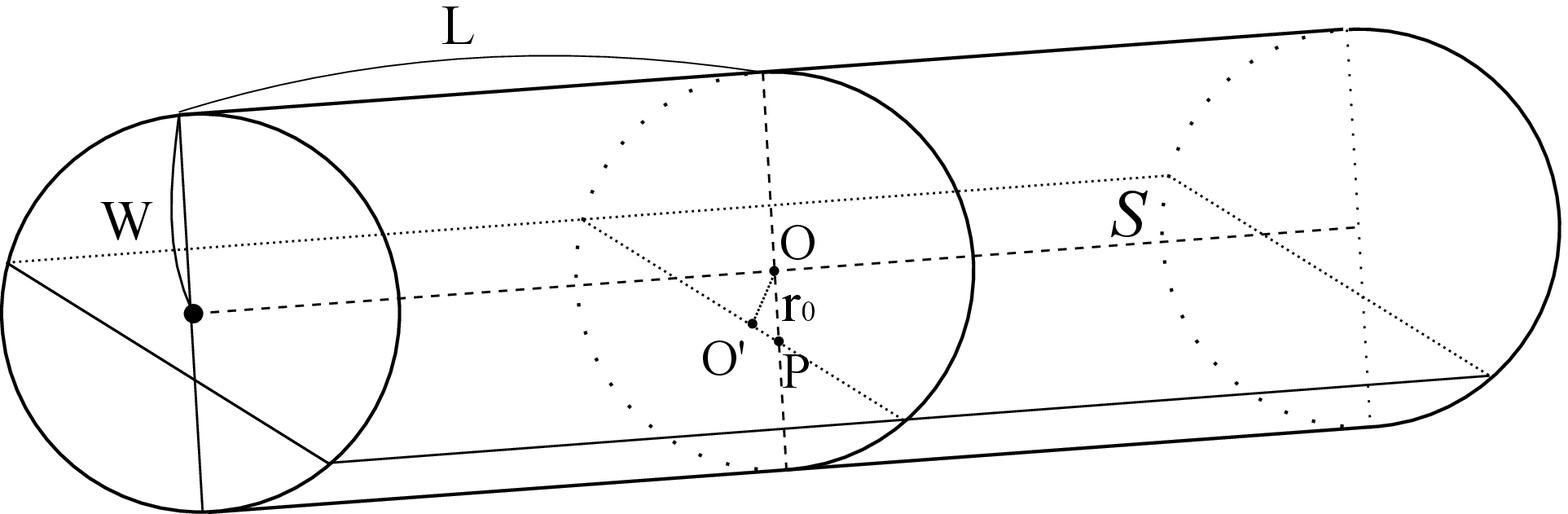}
  \begin{minipage}{0.3\hsize}
    \centering
    \epsscale{1.}
    \plotone{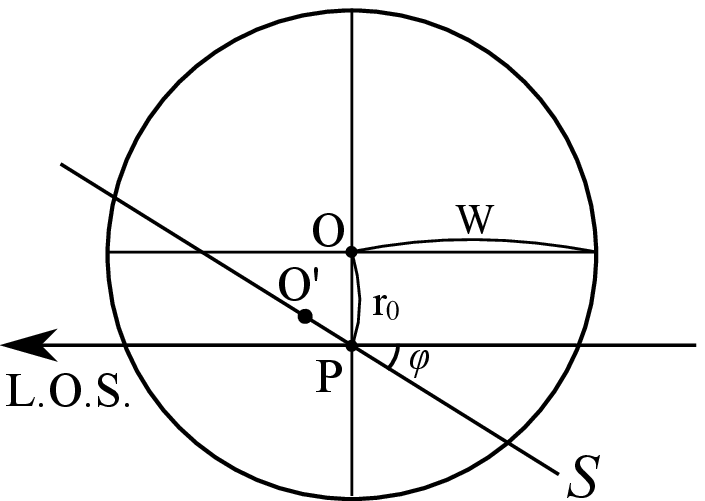}
  \end{minipage}%
  \begin{minipage}{0.7\hsize}
    \centering
    \epsscale{1.}
    \plotone{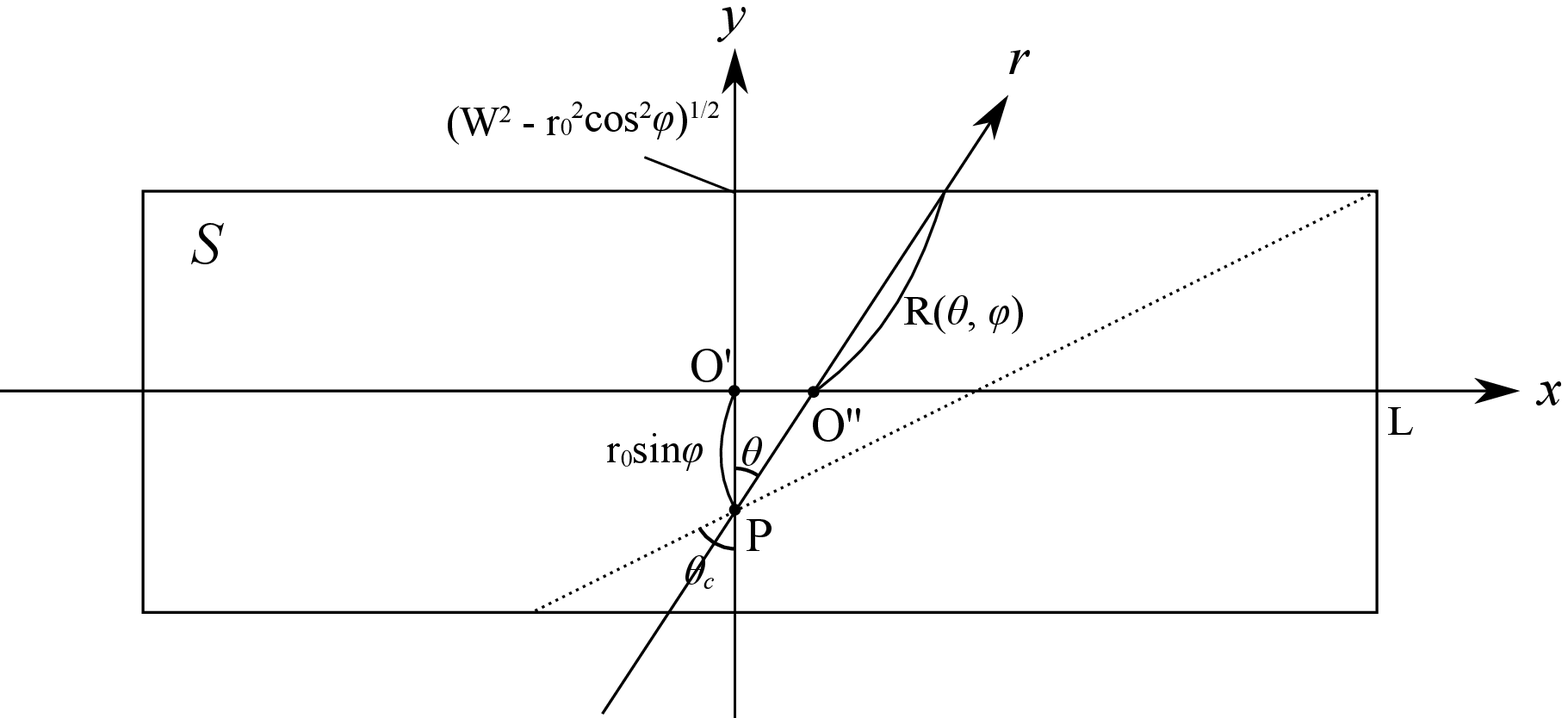}
  \end{minipage}
  \caption{%
    Schematic illustration of a cylindrical model
    to estimate the geometrical effect of a filament on the intensity of FIR radiation.
    The FIR intensity at the position $P$ inside the filament is determined
    by averaging over all directions ($\theta$ and $\phi$).
    We also derived the FIR intensity along the line of sight perpendicular to the cylinder
    crossing the position $P$, as an estimate of the observed intensity.
    \label{fig:appendixB_column}}
\end{figure}
\clearpage
\begin{figure}[!htbp] 
  \centering
  \epsscale{1.0}
  \plotone{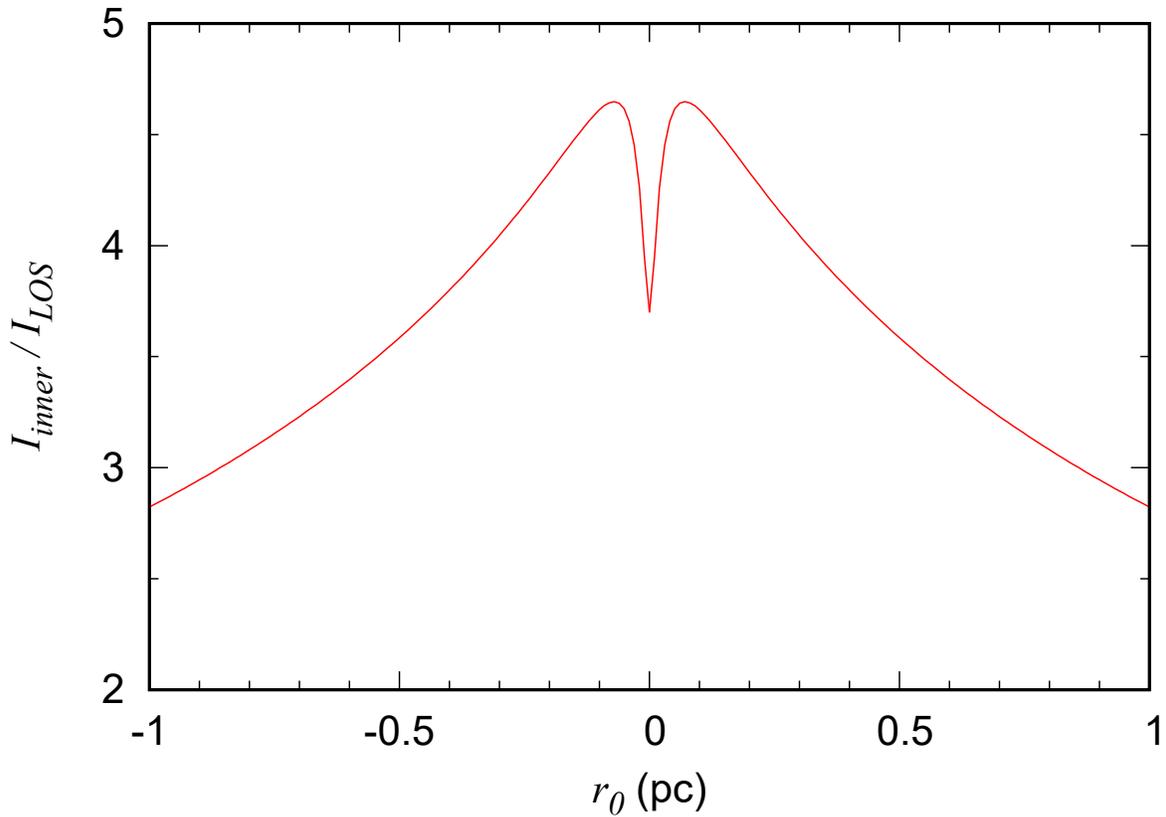}
  \caption{%
    Ratio of the FIR intensity inside the filament
    ($I_{inner}$($\nu$, $r_0$) in Equation ($\ref{eq:appendixB_final}$))
    to
    that along the line of sight perpendicular to the filament
    ($I_{LOS}$($\nu$, $r_0$) in Equation ($\ref{eq:appendixB_los}$))
    as a function of r$_0$.
    Here the Plummer-like density profile (Equation ($\ref{eq:appendixB_plummer}$)) with
    $\rho_c$=9.02$\times$10$^{4}$ cm$^{-3}$, $r_p$=0.012 pc, and $p$=1.84,
    and the cylindrical length of 2.5 pc are assumed.
    The FIR radiation inside the filament within 1 pc can be stronger by a factor of three to four
    compared to the observed value along the line of sight.
    \label{fig:appendixB_Iratio}}
\end{figure}
\clearpage
\begin{figure}[!htbp] 
  \centering
  \epsscale{1.}
  \plotone{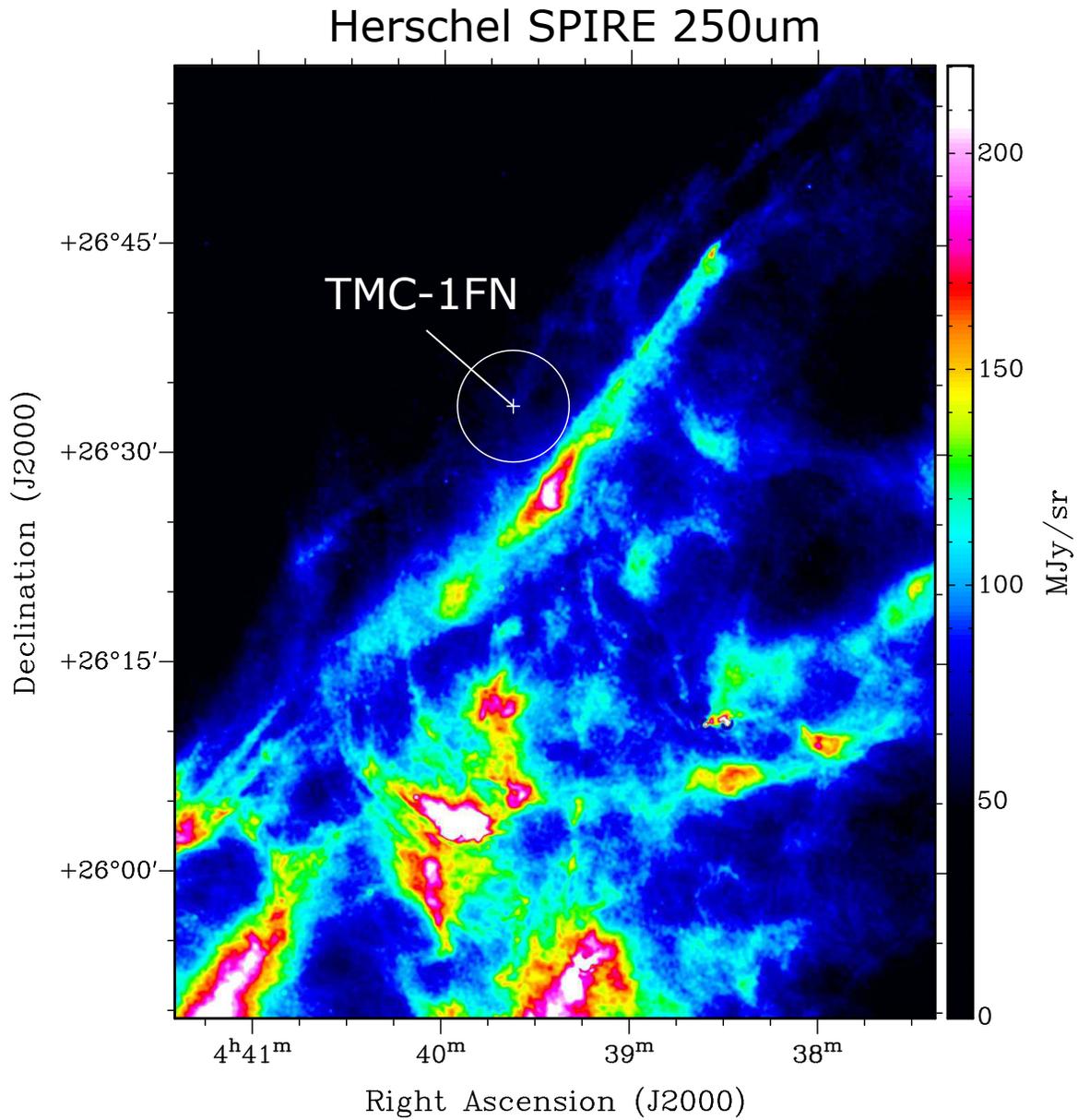}
  \caption{%
    The Herschel SPIRE 250$\mu$m map on the HCL2 region
    obtained from archival data with the observation IDs of 1342202252 and 1342202253.
    White circle represents the observed position of TMC-1FN in the OH 18 cm transition and its beam size.
    A filamentary structure observed in C$^{18}$O (Figure $\ref{fig:HCL2}$) is clearly seen in the southwestern part of TMC-1FN.
  }
  \label{fig:appendixB_250um}
\end{figure}
\clearpage
\appendix
\section{RESULTS OF THE NEW STATISTICAL EQUILIBRIUM CALCULATIONS WITH VARIOUS PARAMETERS}
    Here, we present the results of our statistical equilibrium calculations with different
    OH column densities and H$_2$ ortho-to-para ratios
    (Figures $\ref{fig:SEC_N14_Tk}$--$\ref{fig:SEC_N15_OPR3_nH2}$).
    Assumptions in the calculations for panels (a)--(f)
    are the same as those in Figures $\ref{fig:SEC_N15_Tk}$ and $\ref{fig:SEC_N15_nH2}$.

    Figures $\ref{fig:SEC_N14_Tk}$ and $\ref{fig:SEC_N14_OPR3_Tk}$ show the calculated intensities
    of the OH 18 cm transition as a function of the gas kinetic temperature,
    where the H$_2$ ortho-to-para ratio is assumed to be 0 and 3, respectively.
    The OH column density is assumed to be 3 $\times$ 10$^{14}$ cm$^{-2}$.
    The 1720 MHz line absorption is not reproduced with such a low OH column density,
    as described in Section $\ref{sec:analysis_color}$.
    On the other hand, the effect of the FIR radiation
    is seen for the gas kinetic temperature higher than 30 K,
    since the intensity of the 1612 MHz line in panels (b), (c), (e), and (f) is slightly higher than
    that in panels (a) and (d).
    The 1720 MHz line is also slightly fainter.
    The effect of the line overlaps is negligible
    for the H$_2$ ortho-to-para ratio of 0 (Figure $\ref{fig:SEC_N14_Tk}$),
    whereas the 1667 MHz line is slightly brighter by considering the effect
    of the overlaps for the H$_2$ ortho-to-para ratio of 3 (Figure $\ref{fig:SEC_N14_OPR3_Tk}$).

    Figures $\ref{fig:SEC_N14_nH2}$ and $\ref{fig:SEC_N14_OPR3_nH2}$ show the derived intensities
    of the OH 18 cm transition as a function of the H$_2$ density,
    where the H$_2$ ortho-to-para ratio is assumed to be 0 and 3, respectively.
    Again, the 1720 MHz line absorption is not reproduced
    with the OH column density of 3 $\times$ 10$^{14}$ cm$^{-2}$,
    whereas the effect of the FIR is seen for
    the H$_2$ density lower than $\sim$10$^2$ cm$^{-3}$.

    Figure $\ref{fig:SEC_N15_OPR3_Tk}$ shows the calculated intensities
    of the OH 18 cm transition as a function of the gas kinetic temperature,
    where the H$_2$ density, OH column density, and H$_2$ ortho-to-para ratio are assumed to be
    10$^3$ cm$^{-3}$, 3 $\times$ 10$^{15}$ cm$^{-2}$, and 3, respectively.
    The absorption feature of the 1720 MHz line
    is reproduced for the gas kinetic temperature higher than about 50 K
    by considering the effect of the line overlaps (Figures $\ref{fig:SEC_N15_OPR3_Tk}$ (d)--(f)),
    as shown in Section $\ref{sec:analysis_color}$.
    However, the 1665 MHz line appears in absorption for the gas kinetic temperature
    higher than about 60 K,
    and the 1612 MHz line is too strong compared to the 1665 MHz line
    for the gas kinetic temperature between 50 K and 60 K.

    Figure $\ref{fig:SEC_N15_OPR3_nH2}$ shows the calculated intensities
    of the OH 18 cm transition as a function of the H$_2$ density,
    where the OH column density, the gas kinetic temperature,
    and the H$_2$ ortho-to-para ratio are assumed to be
    3 $\times$ 10$^{15}$ cm$^{-2}$, 20 K, and 3, respectively.
    Qualitative features of the four hyperfine structure lines are similar to
    those derived for the H$_2$ ortho-to-para ratio of 0 (Figure $\ref{fig:SEC_N15_Tk}$).
    On the other hand, the effects of the FIR radiation and line overlaps are smaller
    for the H$_2$ ortho-to-para ratio of 3,
    and the H$_2$ density required to show the 1720 MHz line absorption
    is lower by a factor of $\sim$2--3.
    This can be explained by more efficient collisional excitations to
    the first rotationally excited state levels by collisions with ortho-H$_2$ than para-H$_2$,
    which might compensate for the FIR pumping effect and the effect of the line overlaps.
\section{EFFECTS OF UNCERTAINTIES IN THE COLLISIONAL RATE COEFFICIENTS OF OH}
    We employ the collisional rate coefficients of OH calculated by \citet{Offer1994} in our statistical equilibrium calculations.
    Here, we assess the robustness of our calculations
    by conducting our calculations with the hfs-resolved collisional rate coefficients,
    which are evaluated by applying the M-J random method to the new hfs-unresolved rates reported by \citet{Klos2017}.
    These rate coefficients have been provided by Dr. Fran\c{c}ois Lique.
    Although they are based on the new interaction potential,
    the hfs-resolved collisional rate coefficients are not considered rigorously in the M-J random method \citep{Klos2017}.
    Hence we have just used these rates to confirm our results in the main text.

    Figure $\ref{fig:newrate}$ shows derived intensities of the OH 18 cm transition
    as a function of the H$_2$ density
    with the collisional rate coefficients by \citet{Offer1994} (dotted lines)
    and the new rate coefficients (solid lines).
    It can be seen that all the four solid lines slightly shift to the left
    in comparison with the dotted lines shown in the same color.
    This suggests that the FIR pumping effect is slightly less efficient with the new rate coefficients.
    The H$_2$ density required to show the 1720 MHz line absorption becomes lower
    by a factor of about two
    with the new rate coefficients (Figure $\ref{fig:newrate}$, right).
    Hence, the FIR intensity must be two times stronger
    in order to reproduce the 1720 MHz line absorption with the new rate coefficients,
    assuming the same H$_2$ density.
    Nevertheless, the new rate coefficients do not change the qualitative features
    of the four hfs lines' behavior.
    Furthermore, the absorption in the 1720 MHz line can be reproduced even with the new rate coefficients.
    Therefore, uncertainties in the collisional rate coefficients of OH
    are considered to only have a limited effect, a factor of two at most, on the results discussed in the Section $\ref{sec:Analysis}$.
    Nevertheless, we have to add a caveat that accurate collisional rate coefficients considering
    the hyperfine structure are awaited for a more quantitative analysis
    of the hyperfine anomaly of the OH 18 cm transition.
\section{A GEOMETRICAL EFFECT OF A FILAMENTARY STRUCTURE}
    In this section, we considered the geometrical effect on the intensity of the FIR radiation in TMC-1FN
    by taking into account the density profile and the geometry of a filamentary structure.
    We assume that the filament has a cylinder shape with a radius of $W$ and a length of $2L$,
    as shown in Figure $\ref{fig:appendixB_column}$.
    Its density profile is assumed to be the Plummer-like profile,
    which can be written as a function of the radial distance
    from the central axis of the cylinder ($s$) \citep{Plummer1911}:
    \begin{eqnarray}
    \rho(s) = \frac{\rho_c}{
      \left(
        1 + \frac{s^2}{r_p^2}
      \right)^{\frac{p}{2}}}.
    \label{eq:appendixB_plummer}
    \end{eqnarray}
    By considering a position $P$ inside the filament in Figure $\ref{fig:appendixB_column}$,
    where the distance from the center of the cylinder O is $r_0$,
    the FIR intensity averaged over all directions ($\theta$ and $\phi$) can be written as
    \begin{eqnarray}
    I_{inner}(\nu) =
      u(\nu)
      \frac{2}{\pi} \int_0^{\frac{\pi}{2}} d\phi
      \frac{2}{\pi} \int_0^{\frac{\pi}{2}}d\theta
      \int_{-R(\theta, \phi)}^{R(\theta, \phi)}dr
      \rho(s(r, \theta, \phi)),
      \label{eq:appendixB_I}
    \end{eqnarray}
    where $u(\nu)$ denotes the FIR intensity per hydrogen at a given frequency $\nu$.
    $R(\theta, \phi)$ and $s(r, \theta, \phi)$ are defined as
    \begin{eqnarray}
    R(\theta, \phi) = \begin{cases}
      R_1(\theta, \phi) = \frac{\sqrt{W^2 - r_0^2\cos^2 \phi}}{\cos \theta} &(0 < \theta < \theta_c(\phi))\\
      \label{eq:appendixB_R1}
      R_2(\theta, \phi) = \frac{L}{\sin \theta} &(\theta_c(\phi) < \theta < \frac{\pi}{2})
    \end{cases}
    \end{eqnarray}
    \begin{eqnarray}
    s^2(r, \theta, \phi) = \begin{cases}
      s_1^2(r, \theta, \phi) = s_0^2(r \cos \theta + r_0 \sin \phi)& (0 < \theta < \theta_c(\phi))\\
      s_2^2(r, \theta, \phi) = s_0^2(r \cos \theta) & (\theta_c(\phi) < \theta < \frac{\pi}{2}).
    \end{cases}
    \end{eqnarray}
    Here, $\theta_c(\phi)$ and $s_0^2(x)$ are represented as
    \begin{eqnarray}
    \theta_c(\phi) &=& \tan^{-1}\frac{L}{\sqrt{W^2 - r_0^2\cos^2 \phi}}\\
    s_0^2(x) &=& \left|\left(
      \begin{array}{c}
        0\\
        -r_0\\
      \end{array}
    \right) + x \left(
      \begin{array}{c}
        \cos \phi\\
        \sin \phi\\
      \end{array}
    \right)
    \right|^2\\
    &=& x^2 - 2xr_0\sin \phi + r_0^2.
    \label{eq:appendixB_s02}
    \end{eqnarray}
    We define the following functions
    by substituting the relations $\tilde{r_0}=r_0/r_p$, $\tilde{W}=W/r_p$, and $\tilde{L}=L/r_p$
    for Equations ($\ref{eq:appendixB_R1}$)--($\ref{eq:appendixB_s02}$):
    \begin{eqnarray}
    \tilde{R}(\theta, \phi) = \begin{cases}
      \tilde{R}_1(\theta, \phi) = \frac{\sqrt{\tilde{W}^2 - \tilde{r}_0^2\cos^2 \phi}}{\cos \theta} &(0 < \theta < \theta_c(\phi))\\
      \tilde{R}_2(\theta, \phi) = \frac{\tilde{L}}{\sin \theta} &(\theta_c(\phi) < \theta < \frac{\pi}{2})
    \end{cases}
    \end{eqnarray}
    \begin{eqnarray}
    \tilde{s}_0^2(x) &=& x^2 - 2x\tilde{r}_0\sin \phi + \tilde{r}_0^2\\
    \tilde{s}^2(r, \theta, \phi) &=& \begin{cases}
      \tilde{s}_1^2(r, \theta, \phi) = \tilde{s}_0^2(r \cos \theta + \tilde{r}_0 \sin \phi)& (0 < \theta < \theta_c(\phi))\\
      \tilde{s}_2^2(r, \theta, \phi) = \tilde{s}_0^2(r \cos \theta) & (\theta_c(\phi) < \theta < \frac{\pi}{2}).
    \end{cases}
    \end{eqnarray}
    Then, Equation ($\ref{eq:appendixB_I}$) becomes
    \begin{eqnarray}
    I_{inner}(\nu, r_0) &=& u(\nu) \frac{4r_p\rho_c}{\pi^2}
      \int_0^{\frac{\pi}{2}}d\phi
      \left[
        \int_0^{\theta_c}d\theta
        \int_{-\tilde{R}_1(\theta, \phi)}^{\tilde{R}_1(\theta, \phi)}d\tilde{r}
        \left(1 + \tilde{s}_1^2(\tilde{r}, \theta, \phi)\right)^{-\frac{p}{2}} \right.\\
        &+& \left.
        \int_{\theta_c}^{\frac{\pi}{2}}d\theta
        \int_{-\tilde{R}_2(\theta, \phi)}^{\tilde{R}_2(\theta, \phi)}d\tilde{r}
        \left(1 + s_2^2(\tilde{r}, \theta, \phi)\right)^{-\frac{p}{2}}
      \right].
    \label{eq:appendixB_final}
    \end{eqnarray}
    On the other hand, the FIR intensity along the line of sight perpendicular to the cylindrical axis
    at the position with the distance of $r_0$
    from the center of the cylinder is
    \begin{eqnarray}
    I_{LOS}(\nu, r_0) &=& u(\nu) \int_{-\sqrt{W^2 - r_0^2}}^{\sqrt{W^2 - r_0^2}} \rho\left(\sqrt{r^2 + r_0^2}\right) dr.
    \label{eq:appendixB_los}
    \end{eqnarray}
    Figure $\ref{fig:appendixB_Iratio}$ shows the intensity ratio
    of $I_{inner}$($\nu$, $r_0$) relative to $I_{LOS}$($\nu$, $r_0$)
    as a function of the distance from the center of the cylinder $r_0$.
    Dependence on the FIR frequency ($\nu$) is canceled by the division.
    Here we assume the Plummer-like profile reported by \citet{Malinen2012}
    ($\rho_c$=9.02$\times$10$^{4}$ cm$^{-3}$, $r_p$=0.012 pc, $p$=1.84),
    which is derived from WFCAM extinction map toward the filamentary structure
    harboring TMC-1FN.
    The length of the cylinder is assumed to be 2.5 pc,
    which is estimated from the Herschel 250 $\mu$m map (Figure $\ref{fig:appendixB_250um}$)
    observed toward HCL2 region.
    According to Figure $\ref{fig:appendixB_Iratio}$,
    the FIR intensity at TMC-1FN,
    which is located at $\sim$0.25 pc distant from the center of the filament,
    is estimated to be about 4 times stronger than the observed intensity,
    assuming that the filament is perpendicular to the line of sight direction.
    Therefore, the FIR intensity at 160$\mu$m toward TMC-1FN is estimated to be 300-400 MJy sr$^{-1}$
    according to the observed value with {\it Spitzer} \citep{Flagey2009}.
\listofchanges
\end{document}